\documentclass[acmsmall, nonacm]{acmart}

\AtBeginDocument{%
  \providecommand\BibTeX{{%
    \normalfont B\kern-0.5em{\scshape i\kern-0.25em b}\kern-0.8em\TeX}}}

\usepackage{algorithmic}
\usepackage{graphicx}
\usepackage{textcomp}
\usepackage{xcolor}
\usepackage{subfig}
\usepackage{multirow}
\usepackage{url}
\usepackage{balance}
\usepackage{tabularray}

\begin{document}

\title{``\textit{Filling the Blanks}'': Identifying Micro-activities that Compose Complex Human Activities of Daily Living}

\author{Soumyajit Chatterjee}
\authornote{The author was a research scholar at the Department of Computer Science and Engineering, Indian Institute of Technology Kharagpur, when this work was completed.}
\email{soumyachat@iitkgp.ac.in}
\orcid{0000-0001-5604-2267}
\affiliation{%
  \institution{Nokia Bell Labs, Cambridge}
  \country{United Kingdom}
  \postcode{CB3 0FA}
}

\author{Bivas Mitra}
\affiliation{%
  \institution{Indian Institute of Technology Kharagpur}
  \country{India}
  \postcode{721302}}
\email{bivas@cse.iitkgp.ac.in}

\author{Sandip Chakraborty}
\affiliation{%
  \institution{Indian Institute of Technology Kharagpur}
  \country{India}
  \postcode{721302}}
\email{sandipc@cse.iitkgp.ac.in}

\renewcommand{\shortauthors}{Chatterjee et al.}

\newcommand{\say}[1]{``\textit{#1}''}
\newcommand{\method}{\emph{AmicroN}}
\newcommand{\notesc}[1]{{\textcolor{red}{SC:#1}}}
\begin{abstract}
Complex activities of daily living (ADLs) often consist of multiple micro-activities. When performed sequentially, these micro-activities help the user accomplish the broad macro-activity. Naturally, a deeper understanding of these micro-activities can help develop more sophisticated human activity recognition (HAR) models and add explainability to their inferred conclusions. Previous research has attempted to achieve this by utilizing fine-grained annotated data that provided the required supervision and rules for associating the micro-activities to identify the macro-activity. However, this ``bottom-up'' approach is unrealistic as getting such high-quality, fine-grained annotated sensor datasets is challenging, costly, and time-consuming. Understanding this, in this paper, we develop \method{}, which adapts a ``top-down'' approach by exploiting coarse-grained annotated data to expand the macro-activities into their constituent micro-activities without any external supervision. In the backend, \method{} uses \textit{unsupervised} change-point detection to search for the micro-activity boundaries across a complex ADL. Then, it applies a \textit{generalized zero-shot} approach to characterize it. We evaluate \method{} on two real-life publicly available datasets and observe that \method{} can identify the micro-activities with micro F\textsubscript{1}-score $>0.75$ for both datasets. Additionally, we also perform an initial proof-of-concept on leveraging the state-of-the-art (SOTA) large language models (LLMs) with attribute embeddings predicted by \method{} to enhance further the explainability surrounding the detection of micro-activities.
\end{abstract}





\maketitle

\section{Introduction}
\label{intro}
\begin{figure*}
	\centering
	\subfloat[``Bottom-up'' Approach\label{fig:bottom_up}]{
		\includegraphics[width=0.44\textwidth, keepaspectratio]{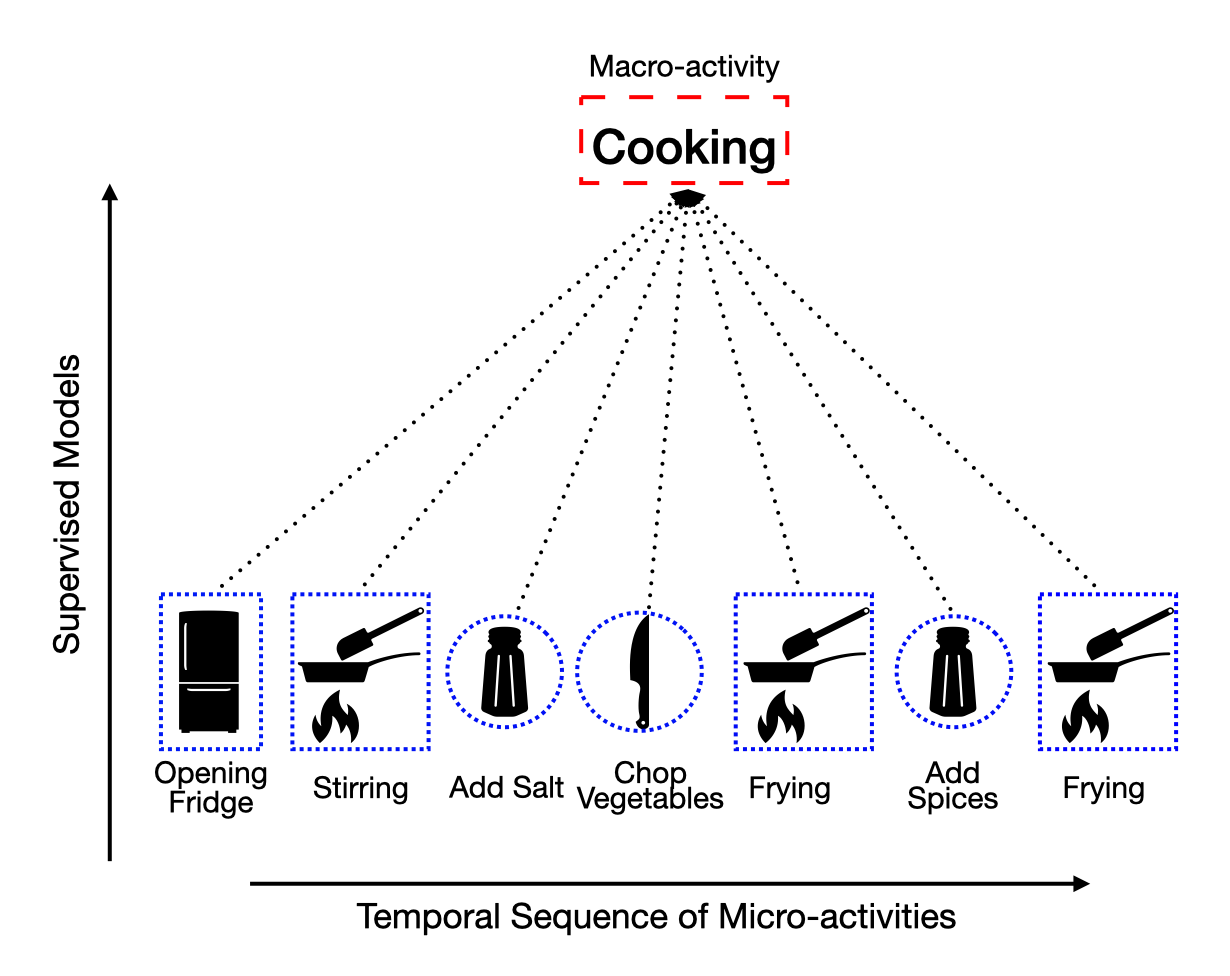}
	}
	\subfloat[Proposed ``Top-down'' Approach\label{fig:amicron_top_down}]{
		\includegraphics[width=0.56\textwidth, keepaspectratio]{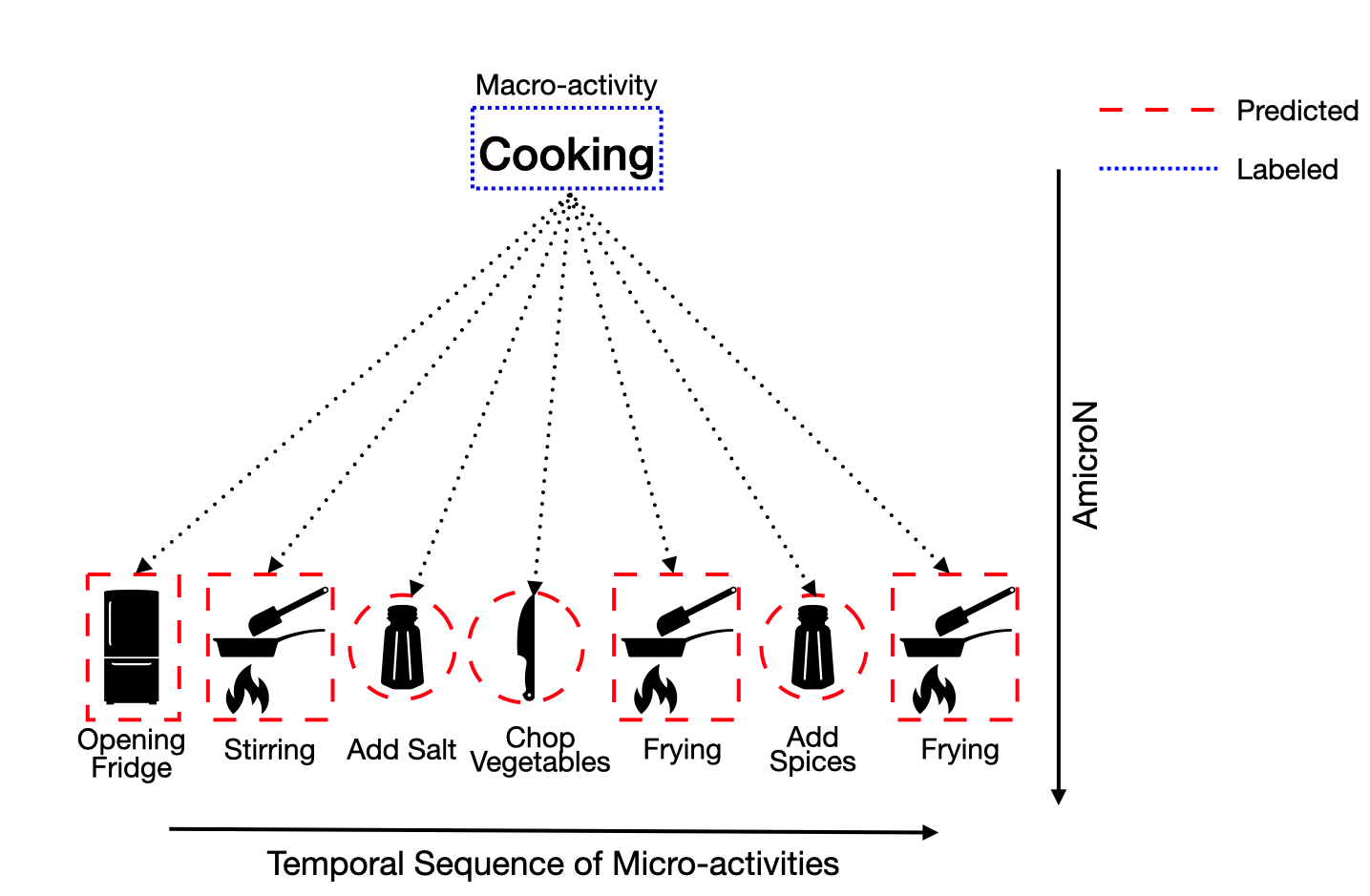}
	}	
	\caption{(a) Conventional bottom-up approach of using granular micro-activity labeled data to identify the macro-activity label vs (b) our proposed top-down approach, named \method, which decomposes a coarse-grain macro-activity labeled data and identifies the hidden fine-grained micro-activities.}
	\label{fig:key_diffs}
\end{figure*}

\textbf{Motivation:} Since the advent of different smart wearables and other pervasive sensing devices, one of the core research drives has been to identify and characterize sophisticated human activities of daily living (ADLs). However, complex ADLs are often amalgamations of multiple micro-activities~\cite{micro} that, when performed synergistically, complete the primary task posed by a \textit{macro-activity}. For example, as shown in \figurename~\ref{fig:key_diffs}, one of the complex macro-activities can be \say{cooking}, which in turn, is composed by several micro-activities like \say{chopping vegetables}, \say{stirring}, \say{adding spices}, etc. Undoubtedly, precise identification of these micro-activities can provide a more explainable understanding of the composition of these ADLs~\cite{10.1145/3580804}, which in turn can help to develop more sophisticated human activity recognition (HAR) models that can be used to provide a variety of specialized services. For example, identifying such fine-grained micro-activities can help us to isolate the critical points of failure in industrial applications, which is one of the major focus areas of future smart industry applications~\cite{s20010109,LIU2017295,Liu_Cheng_Liu_Jia_Rosenblum_2016}. Similarly, fine-grained micro-activity information can be used to develop sophisticated context-aware applications for activities performed in environments like a smart home. For example, applications like monitoring the cognitive health of the users from activity patterns~\cite{7299406,mohammad2017dataset,franklin2021designing,new_article_1} have already shown their potential under such context. Understanding these potential avenues, we, in this paper, consider designing a HAR system that provides a microscopic view of the complex ADLs which in turn can help such a system become more explainable.

\noindent
\textbf{Primary Challenge:} However, obtaining annotations for sensor data is highly costly~\cite{10.1145/3626960} and time-consuming, and the added requirement of getting granular micro-activity\footnote{\textbf{Key Terminologies:} We consider inherently atomic activities as \textit{micro-activities}. Therefore, both of these terms (\textit{atomic activity} and \textit{micro-activity}) are used interchangeably throughout this draft. On the other hand, \textit{macro-activities} are activities that consist of more than one micro-activities performed in a specific temporal sequence.} annotations can make this process even more challenging. Typically, the sensor data annotation process involves human annotators labeling the sensor data using information from alternate modalities, like videos or audio captured from the smart environment. Usually, these annotations are shallow, encompassing only a few basic macro-activities. This is because the annotators often fail to identify different fine-grained activity details~\cite{fleury2009svm,roggen2010collecting,cmummac,conflict} subjected to their individual biases and understanding (see Section~\ref{motivation}). 
Naturally, with the availability of granular annotations being the primary bottleneck, works like~\cite{junker2008gesture,saguna2013complex,LIU2017295,7299406,roy2016ambient,LIU201641,Liu_Cheng_Liu_Jia_Rosenblum_2016,10.1145/3580804}, which rely on characterizing complex ADLs by training \emph{supervised} ML-models with granular micro-activity labeled data in a ``bottom-up'' manner (see \figurename~\ref{fig:bottom_up}), becomes entirely unrealistic.

\noindent
\textbf{Opportunities:} Interestingly, even across different complex ADLs, there are certain activities that are inherently \textit{atomic} in nature. For example, activities like \say{open cupboard} and \say{take spoon} are inherently atomic.
Regardless of the level of granularity of the annotations in the dataset, such atomic macro-activity labels actually present us with an opportunity to use the locomotive patterns identified in these activities to expand a more complex macro-activity like \say{take baking pan} where the subject first performs the micro-activity of \say{opening the cupboard} and then \say{taking out the utensil}. Additionally, complex ADLs are often composed of different micro-activities that involve significantly different movement patterns or gestures, which, if properly segmented~\cite{LIU201641} in time, can provide us with unique patterns for recognizing them separately.

\noindent
\textbf{Problem Formulation:} Given these opportunities, we formulate our problem statement as follows. Consider data from one or more bodyworn inertial sensors $\mathcal{I}^{t}_{t+1}$, with the timestamped macro-activity label $\mathbb{L}^{t}_{t+1}$ collected for the time duration $[t, t+1]$. We in this paper, aim to develop the framework \method, that can identify the sequence of micro-activities $\mathbb{M}^{t}_{t+1} = {m_1, m_2, \hdots, m_n}$ that compose the macro-activity defined by $\mathbb{L}^{t}_{t+1}$. Notably, unlike the aforementioned bottom-up approaches, we do not require access to granular micro-activity annotations and utilize the existing shallower macro-activity annotations already available in the dataset. More specifically, \method{} first searches for the atomic macro-activities within the dataset using a time-thresholding-based approach and then uses this information available for identifying the hidden micro-activities in a top-down manner (see \figurename~\ref{fig:amicron_top_down}).

\noindent
\textbf{Our Contribution:} However, regardless of the available annotation granularity, identifying micro-activities is not straightforward, especially when no prior information regarding macro-activity composition is available. Nevertheless, as these micro-activities often produce entirely different locomotive signatures, a judicious solution could be to first segment them temporally. Based on this broad intuition, \method{} starts by performing an \emph{unsupervised change-point detection} to precisely detect the activity boundaries, which can give the framework an idea regarding the number of micro-activities and their transitions.

Although this change-point-based segmentation approach allows \method{} to identify the micro-activity boundaries without any external supervision, the exact recognition of the micro-activity still remains a challenge as the atomic macro-activity annotations can give us access to only a limited set of activities. This, in turn, restricts the usage of supervised models for identifying the micro-activities as there can be many other activities that may not have been observed in the limited training dataset containing the atomic macro-activity labels only. \method{} resolves this challenge by incorporating \emph{generalized zero-shot learning} for predicting the micro-activities in this restricted setup.

Typically, zero-shot learning for HAR leverages some form of semantic representation of the \emph{action verbs} (like \say{chopping}) to train the model~\cite{tong2021zero}. However, the performance of zero-shot models is known to be heavily impacted by the choice of this semantic space~\cite{tong2021zero,al2020zero}. Furthermore, multiple IMU units can be attached to different body parts of the subject in an unrestricted smart environment that may provide input to the framework. Subsequently, an incorrect selection of sensors during the training and prediction phases can mislead the model significantly~\cite{keally2011pbn,zhang2013senstrack}. In this paper, we mitigate these two modeling concerns as follows.

We resolve the problem regarding the choice of semantic space by using latent embeddings for the actions performed in the environment. More specifically, in this paper, we try two different types of semantic representations, namely -- (a) \emph{verb attribute}~\cite{Zellers2017ZeroShotAR} and (b) motion-based semantic attributes~\cite{lara} as latent representations. Similarly, to mitigate the concern regarding the sensor selection, we apply state-of-the-art (SOTA) unsupervised dimensionality reduction techniques like uniform manifold approximation and projection for dimensionality reduction (UMAP)~\cite{mcinnes2018umap-software, SMG2020}, during the training and prediction phases to assess the maximum variations caused by the sensor units, which are explicitly responsible for capturing the required body movements. In summary, the primary contributions of this paper are as follows.
\begin{enumerate}
    \item We develop \method{} which can decompose complex ADLs into their constituent micro-activities \textbf{without} the requirement of having \textbf{access to a vast training dataset} containing granular micro-activity annotations. To the best of our knowledge, this is the \textbf{first approach} that addresses this problem in a \textbf{``top-down manner''}, unlike the conventional bottom-up approach of identifying the final macro-activity from its constituent micro-activities.
    \item We systematically evaluate \method{} on two public datasets -- one capturing the kitchen~\cite{cmummac} activities and the other capturing warehouse~\cite{lara} activities. We observe that \method{} can indeed detect the micro-activity boundaries and perform appreciably well in identifying the micro-activities using an intelligently designed pipeline with unsupervised change-point detection followed by generalized zero-shot learning.
    \item Finally, as a qualitative analysis, we briefly show the potential of \method{} to work as a plug-and-play module over the SOTA large language models (LLMs) for generating micro-activity annotations.
\end{enumerate}

The remainder of the paper is arranged as follows. First, we summarize the recent state-of-the-art (SOTA) approaches and highlight the key takeaways (Section~\ref{rel_work}). Following the literature survey, we next describe in detail the datasets (Section Section~\ref{datasets}) that we use in this paper to perform motivational experiments (Section \ref{motivation}) and evaluate \method. We next design \method{} (Section \ref{overview}) and describe in detail the different steps involved in training (Section \ref{train}) as well as prediction (Section \ref{pred}) phases of \method. Next, we briefly describe the overall setup (Section \ref{grnd_annotation} and Section \ref{setup}) that is used to evaluate \method{} in a principled manner. Finally, we perform a rigorous evaluation of \method{} (Section \ref{eval}) and conclude the paper with some future directions (Section \ref{discussion}).
\section{Related Work and Preliminaries}
\label{rel_work}
This section summarizes the literature survey that we have performed surrounding the problem of identifying micro-activities. To summarize this systematically, we first discuss the works that followed the bottom-up approach, which is entirely opposite to what \method{} aims to achieve; we then discuss the critical limitations of this bottom-up approach and highlight the challenges associated with obtaining granular ground truth annotations for micro-activities. Finally, this section also briefly discusses zero-shot learning for HAR.
\subsection{The Bottom-up Approach: Stitching Micro-activities to Identify Macro-activities}
In the past there has been a plethora of works like~\cite{junker2008gesture,roy2016ambient,Liu_Cheng_Liu_Jia_Rosenblum_2016,LIU201641,LIU20161015,LIU2017295,10.1007/978-3-030-67667-4_21} that have discussed the idea of identifying a complex macro-activity by stitching the micro-activities modeled using a supervised approach. For example, works like~\cite{junker2008gesture,Liu_Cheng_Liu_Jia_Rosenblum_2016,LIU201641,LIU2017295,LIU20161015,10.1007/978-3-030-67667-4_21} define a set of atomic activities that could easily be detected using wearable sensors. For example, \cite{LIU201641} defined a pattern dictionary containing \textit{shapelets}, which are essentially specific time-series signatures for fine-grained atomic activities. Similarly, \cite{Liu_Cheng_Liu_Jia_Rosenblum_2016} used a specially designed \textit{record of triplets}, which contained sequences of atomic activities along with their specific start and end times. Following this general approach of defining a training dataset with fine-grained atomic activity details, most of these works then used supervised models like Hidden Markov Models~\cite{junker2008gesture,LIU20161015}, probabilistic models~\cite{Liu_Cheng_Liu_Jia_Rosenblum_2016, LIU2017295}, Conditional Random Fields~\cite{10.1007/978-3-030-67667-4_21}, and occasionally also relied on causal rules~\cite{10.1007/978-3-030-67667-4_21} to finally thread all the atomic (or micro)-activities to identify the complex macro-activity. This approach allowed these works to explore different problem areas ranging from detecting complex ADLs and instrumental ADLs (or IADLs)~\cite{Liu_Cheng_Liu_Jia_Rosenblum_2016, LIU201641}, assessing restaurant processes~\cite{LIU2017295} to detecting offensive play in a game of basketball~\cite{10.1007/978-3-030-67667-4_21}. Also, there have been works like~\cite{roy2016ambient} which further extended these approaches and applied them to scenarios with multiple inhabitants in the same smart space. More recently, works like~\cite{10.1145/3580804} have revisited the usage of such micro-activity description as \textit{concept sequences} for adding explainability to the recognition of broad macro-activities.

\noindent
\textbf{Key limitations:} Although all of these aforementioned work helped in recognizing complex ADLs, this \textbf{general assumption} that the \textbf{micro-activities are labeled apriori} and are stitched together in a temporal sequence to \textbf{identify the macro-activity} is \textbf{problematic}. A major \textbf{pitfall} of this bottom-up design is that it requires a \textbf{significant amount of annotation effort}. For example, it is much easier for an annotator to mark a longer duration with a coarse-grain label \say{cooking} rather than providing fine-grained labels like \say{taking utensils}, \say{washing dishes}, \say{chopping}, etc. This becomes even more \textbf{challenging} if the individual \textbf{micro-activities} occur in quick successions with \textbf{lesser transition time} between them. It has already been observed that these cases often create \textbf{low-quality datasets} that have a significant amount of \textbf{label jitter} present in them~\cite{zhang2020syncwise}.
\subsection{Annotations and Micro-Activities}
Following the aforementioned discussion, we summarize the key challenges observed in obtaining annotations for HAR applications. Typically, annotations for HAR have been obtained from human-in-the-loop-based solutions. However, many recent works like~\cite{conflict,active1,chatterjee2020laso} have already pointed out the limitations, like the problem of missing and conflicting labels prevalent in this approach. However, other than these problems, another major concern that often plagues human-in-the-loop annotations is the granularity of annotations. Although more straightforward physical tasks like running, walking, etc., may not need further granular annotations, complex tasks often contain several fine-grained micro-activities~\cite{mohammad2017dataset}. Interestingly, papers like~\cite{cmummac} have observed the ambiguity surrounding the annotation process for complex activities. For example, the kitchen activities in~\cite{cmummac} could have been annotated in three different ways -- with the recipe (more superficial), action, or gesture (most fine-grained) level granularity. Additionally, with unrestricted environments, like in~\cite{cmummac}, people often multitask and perform different activities together, further complicating the process of generating an unambiguous label.

Notably, a standard solution to this ambiguity has been to generate less complicated and often shallow annotations to reduce overall variability~\cite{cmummac}. For example, in~\cite{cmummac}, the authors prefer a recipe-level annotation instead of the other two granular variations. However, this reduces the granularity of the overall annotations. Although machine-assisted solutions like~\cite{active1,van2017experience} can allow obtaining fine-grained labels, end-user involvement may be challenging if the smart environment is designed for the aged and the differently-abled population. Also, completely automated solutions for generating annotations like~\cite{chatterjee2020laso,xing2018enabling} will be restricted in terms of granularity, depending on the choice of their auxiliary modalities.
\subsection{Zero-Shot Activity Prediction}
Observing the challenges in gathering annotations, there has been significant improvement in learning approaches for continuous sensing systems. Instead of direct supervision, many works have also started looking into strategies that can work even when labeled datasets are unavailable. Out of these approaches, zero-shot learning has been the most prominent, where the system can predict labels even when the learning model does not have access to all the labels present in the test set~\cite{larochelle2008zero}. Typically, for human activity recognition from non-visual sensing approaches, the conventional way of generating zero-shot labels is using the word-level embeddings using state-of-the-art models like Word2Vec~\cite{mikolov2013efficient} and GloVe~\cite{pennington2014glove} of the annotated labels~\cite{al2020zero,matsuki2019characterizing}.

Notably, many recent works like~\cite{ wang2017zero,tong2021zero} have shown that these automatically generated word embeddings perform poorly compared to handcrafted and video-based embeddings. Interestingly, several works have also explored the chances of using verb attribute vectors that defined body movement patterns~\cite{xian2016latent,cheng2013towards, rueda2018learning}. Although defining such attributes is more straightforward for more uncomplicated mobility-based physical activities, defining such granular verb attributes with complex ADL(s) becomes very complicated. Understanding these concerns, some of the other works like~\cite{Zellers2017ZeroShotAR} have introduced verb attributes that capture more word embedding-based properties like transitivity of the verb along with some generic action and gesture-related attributes, which makes them less complicated to design and more robust for zero-shot learning.

In addition to these concerns, another major limitation of zero-shot learning has been the problem of hubness. This problem of \textit{hubness} causes many unnecessary verb attribute embeddings appearing as ``universal'' nearest neighbors~\cite{dinu2014improving,radovanovic2010hubs}, which in turn limit the practical applicability of the zero-shot model to some extent. Additionally, with massive corpora used for building these embedding models, many unrelated words may appear more similar to the action, which creates significant confusion while verifying the end results~\cite{al2020zero}.
\section{Datasets}
\label{datasets}
We first look into some well-known publicly available datasets that capture complex ADLs to understand the critical limitations in obtaining annotations for labeling sensor signals. This section discusses the datasets and their annotation procedure in detail. We also discuss how the different parts of this paper will use these datasets to draw motivations and evaluate the proposed methodology in a principled manner. The details follow.
\subsection{Kitchen Dataset}
The kitchen dataset~\cite{cmummac} contains the data of different subjects performing various complex kitchen activities while cooking predefined dishes like a brownie, sandwiches, eggs, pizzas, salads, and sandwiches. In our paper, we consider a part of the entire dataset with $12$ subjects for whom the annotations are present\footnote{\url{http://kitchen.cs.cmu.edu/main.php} (Accessed: \today)}. Notably, these annotations were provided by a dedicated annotator using the first-person videos as the ground-truth modality using a specifically designed annotation tool~\cite{cmummac}. Notably, the annotator was explicitly asked to form the labels from a predefined corpus of verbs, prepositions, and nouns in a structured way of constructing the labels with recipe-level granularity~\cite{cmummac}.
Concerning the sensing modalities, the subjects are tracked through multiple time-synchronized body-worn sensors, including $5$ wired IMU sensors attached to the subject's arms, legs, and back. There are also sensors like microphones, motion capture, RFID(s), and e-watches to record the activity signatures of the subjects in the kitchen. Considering the input from the $5$ wired IMU sensors (sampled at $125$Hz), although time-synchronized, these sensors' polling is asynchronous. Therefore, we first preprocess the data to achieve a synchronous sensor polling using a moving window averaging approach with a window length of $10$ms. This approach synchronizes all the IMU data obtained from the $5$ wired IMU units, albeit this reduces the sampling rate of the preprocessed data to $100$Hz.
\begin{table}[]
	\scriptsize
	\centering
	\caption{Semantic Attributes in LARa Dataset. The coarse-grain activity label for these semantic attributes is \say{Standing}}
	\label{tbl:semantic_attributes}
	\begin{tabular}{|c|c|c|c|c|c|c|c|c|}
\hline
\textbf{Gait Cycle} &
  \textbf{Step} &
  \textbf{\begin{tabular}[c]{@{}c@{}}Standing\\ Still\end{tabular}} &
  \textbf{Upwards} &
  \textbf{Centred} &
  \textbf{Downwards} &
  \textbf{\begin{tabular}[c]{@{}c@{}}No Intentional\\ Motion\end{tabular}} &
  \textbf{\begin{tabular}[c]{@{}c@{}}Torso\\ Rotation\end{tabular}} &
  \textbf{Right} \\ \hline
0             & 0                & 1                   & 0                   & 0                    & 0             & 1                 & 0                & 0             \\ \hline
\textbf{Left} & \textbf{No Arms} & \textbf{Bulky Unit} & \textbf{Handy Unit} & \textbf{Utility Aux} & \textbf{Cart} & \textbf{Computer} & \textbf{No Item} & \textbf{None} \\ \hline
0             & 1                & 0                   & 0                   & 0                    & 0             & 0                 & 1                & 0             \\ \hline
\end{tabular}
\end{table}
\subsection{LARa Dataset}
\label{lara_dataset}
The LARa (\textbf{L}ogistic \textbf{A}ctivity \textbf{R}ecognition Ch\textbf{a}llenge) dataset~\cite{niemann2020lara,lara} contains the data of $15$ subjects in a warehouse environment focusing on $7$ coarse-grain logistic activities like standing, walking, etc. The coarse-grain label space also contained the label `None' to include the irrelevant activities. In addition to these coarse-grain labels, the dataset also includes $18$ semantic attributes like gait cycle, torso rotation, etc., providing fine-grained information regarding the movement patterns of different body parts while performing these broad activities (an example is shown in \tablename~\ref{tbl:semantic_attributes}). Interestingly, this entire dataset is annotated using the optical marker-based motion capture (or OMoCap), further supported by RGB videos of the environment with manual annotations. Concerning the sensor data, this dataset also contains sensor data from $5$ IMU sensors (sampled at $100$Hz) attached to different body parts\footnote{\url{https://zenodo.org/record/3862782} (Accessed: \today)}.
\subsection{Health Smart-Home Dataset}
The health smart-home dataset (HSD)\footnote{\url{https://lig-getalp.imag.fr/health-smart-home-his-datasets-2/} (Accessed: \today)} contains the data from $15$ subjects~\cite{fleury2009svm}, where each participant was asked to live alone in a flat and perform some of the predefined ADL(s). The flat is completely equipped, and several sensors, including microphones, infrared sensors, cameras (not in the bathroom), hygrometers, and temperature sensors, are installed to sense different activities the subject performs. A single IMU sensor placed under the subject's armpit is used to capture the locomotion. The dataset is also annotated with $7$ broad-level activities using the ground-truth input from $5$ different video cameras installed in the setup. These activities were predefined, and the subject was explicitly instructed to perform all of these, albeit no restrictions were imposed on the subject regarding their order and duration. These $7$ activities included most ADL(s) like sleeping, feeding, communication, hygiene, resting, toilets, and communication. Notably, many different tasks were clubbed under these $7$ broad-level categories. For example, the entire activity labeled as \say{Feeding} was described as -- \say{you prepare breakfast with the equipment and the ingredients in the kitchen cupboard and then eat (using the kitchen table). Then, you wash the dishes in the kitchen sink.}
\subsection{Dataset Usage Description}
Although all these datasets contain the necessary information for further analysis, the dataset needs to have the annotations done at least to a certain degree of granularity, making it suitable for us to evaluate \method. For example, as mentioned above, HSD provides a detailed activity pattern from a smart home environment. However, the dataset is very coarsely labeled. Although it cannot be used to evaluate \method{}, it provides us with observations to highlight why getting granular annotations is challenging and how sensor signals can allow us to investigate further.
\section{Motivational Experiment}
\label{motivation}
Before developing \method, we first start with a set of experiments to understand why the designed top-down approach is more realistic than the usual bottom-up approach discussed in Section \ref{rel_work}. To do this analysis, we first perform a set of experiments on some of the publicly available ADL datasets discussed in Section \ref{datasets}. Subsequently, we highlight the key observations and revisit the advantages that motivate the development of the \method.
\subsection{Pilot Study}
\begin{figure*}[]
	\centering
	\captionsetup[subfigure]{}
	\begin{center}
		\subfloat[\label{fig:grenoble_rf}]{
			\includegraphics[width=0.50\linewidth,keepaspectratio]{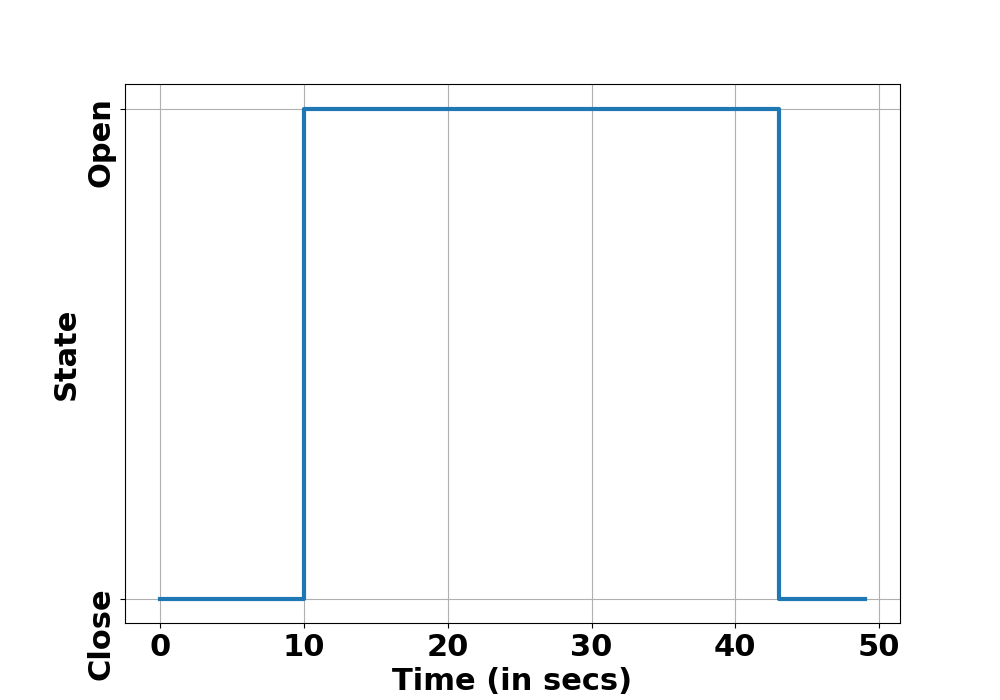}
		}
		\subfloat[\label{fig:grenoble_imu_dress}]{
			\includegraphics[width=0.50\linewidth,keepaspectratio]{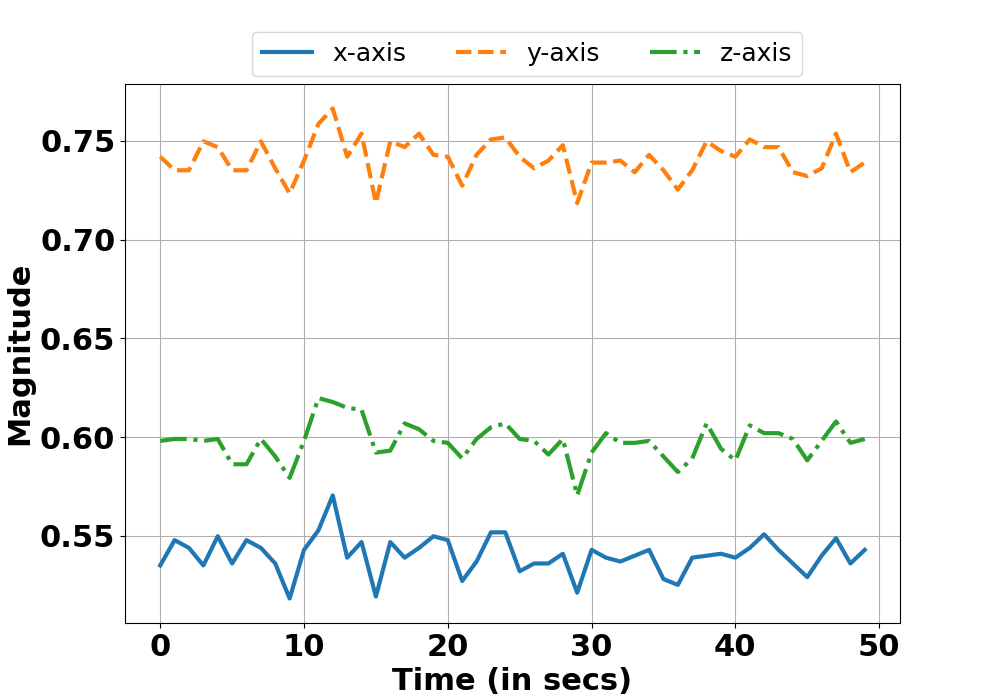}
		}
	\end{center}
	\caption{Issues with coarse-grain labeling in the HS dataset -- Changes in (a) furniture sensor for the activity class and (b) raw actimetry data for the activity class \say{Dressing/Undressing}.}
	\label{fig:grenoble_change}
\end{figure*}
We first start by analyzing the HSD to assess the quality of its given annotations. From initial observation, we find that the entire set of activities performed by the subject in the smart home is covered by only $7$ \textbf{coarse-grain activity labels}. Although the \textbf{labels are accurate and consistent} in terms of representing the broad activity performed by the subject, the annotators \textbf{conveniently ignore multiple micro-activities} that compose those $7$ complex activities. For example, in the case of the activity label \say{Dressing/Undressing}, there are multiple associated micro activities, including opening and closing of drawer, folding and arranging the clothes, etc., which have very distinct signatures on physical sensors. For example, from \figurename~\ref{fig:grenoble_rf} and \figurename~\ref{fig:grenoble_imu_dress}, we observe that both the bodyworn inertial as well as the furniture sensors change, indicating that the subject opened a drawer and then closed it, albeit the label only defines the activity as \say{Dressing/Undressing} only.
\begin{figure*}[]
	\centering
	\captionsetup[subfigure]{}
	\begin{center}
		\subfloat[\label{fig:cmu_imu1_change}]{
			\includegraphics[width=0.50\linewidth,keepaspectratio]{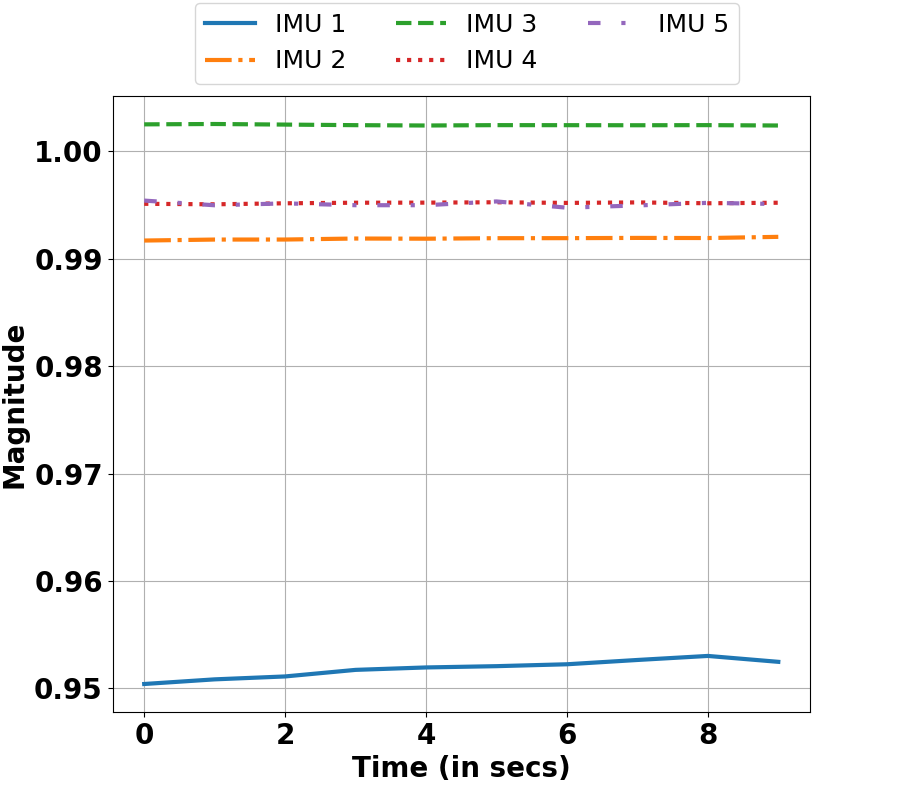}
		}
		\subfloat[\label{fig:cmu_imu2_change}]{
			\includegraphics[width=0.50\linewidth,keepaspectratio]{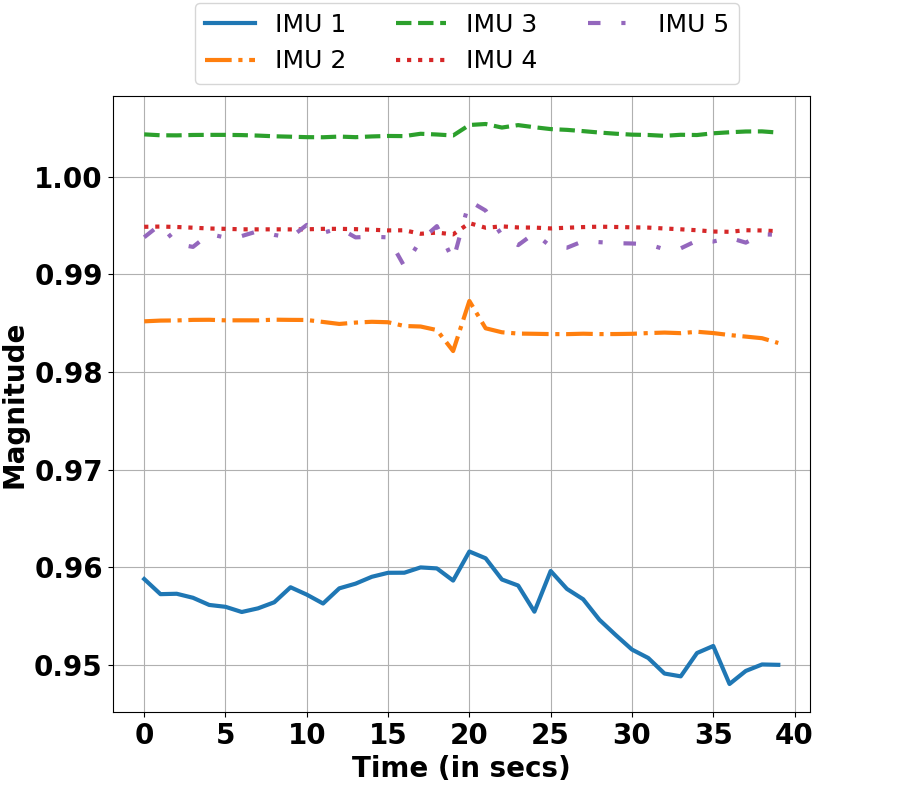}
		}
	\end{center}
	\caption{Issues of subjectivity in the annotations of the kitchen dataset -- Variation of the accelerometer signatures for the subjects (b) S11 and (c) S06 for the activity label ``pour big bowl into baking pan.''}
	\label{fig:cmu_changes}
	\Description{This figure depicts the issue of subjectivity in the choice of defining the activity boundaries for the same activity of \say{pouring big bowl into baking pan} from the kitchen dataset. Subfigures show the corresponding IMU signatures (magnitude) for the users (a) S11 and (b) S06, respectively.}
\end{figure*}

Following these observations, we move to the kitchen dataset, which, compared to the HSD dataset, has labels with \textbf{higher granularity}. However, a deeper investigation of the dataset reveals a significant \textbf{lack in consistency} concerning the granularity across the labels for all the subjects. For example, considering subjects S22 and S16 from the kitchen dataset, we observe from the ground truth video data that both the subjects scrape the big bowl with a fork for the activity \say{pour big bowl into baking pan}. However, for subject S22, the annotator ignores that and includes it as a part of the invalid label \say{None}. Whereas, for subject S16, the annotator considered it to be a part of the broad label \say{pour big bowl into baking pan}, which generates a completely different IMU signature (compare \figurename~\ref{fig:cmu_imu1_change} and~\ref{fig:cmu_imu2_change}). 

Similarly, for subject S07, we observe that the label space contains the detailed label \say{open cupboard bottom right}; however, when the subject opens the bottom cupboard to \say{take baking pan}, the activity \say{open cupboard bottom} is ignored and considered as a part of the broad activity label \say{take a baking pan.} Thus, from the perspective of locomotive signatures captured from the subject, the primary actions are absent from the annotated label even though an already recognized action was performed. Interestingly, the authors of the kitchen dataset also faced a similar dilemma during the annotation process, as highlighted in~\cite{cmummac}. They found that the same activity can be annotated from different perspectives, ranging from a recipe level (like \say{break one egg}) to a fine-grained movement level (like \say{reach forward with left hand}) during the annotation process. However, they observed that any specific choice or idea, with higher granularity, can result in more ambiguity. For example, they observed that when the subjects perform complex activities, like \say{pouring in brownie mix} while \say{stirring the bowl}, it becomes extremely difficult to annotate at an extremely higher granularity (based on movement or gesture) the action due to the high chances of ambiguity in such actions. Following this observation, they selected a set of $29$ recipe-level labels with limited granularity and lesser ambiguity.

\subsection{Key Observations}
We summarise the key observations gained from the aforementioned motivational experiment as follows.
\begin{enumerate}
    \item \textbf{Obtaining granular micro-activity annotations is challenging:} One of the key observations that we gain from the motivational study is that obtaining granular micro-activity annotations for practical cases is not straightforward. Naturally, a general tendency to make an \textbf{annotation} process \textbf{less tedious and straightforward} is by accepting \textbf{shallower coarse-grain labels} that represent the broad activities getting performed.
    \item \textbf{Higher granularity may lead to more erroneous and inconsistent labeling:} Notably, an important factor while getting annotations is also the broad task description, i.e., whether the task clearly specifies the requirement for more granular labels from the annotator. However, we observe that with a \textbf{higher degree of granularity}, there are also \textbf{higher chances of getting more inconsistent and even erroneous labels}. Additionally, granular micro-activity annotations are also prone to \textit{label jitter} where the activities may not be accurately mapped to their individual time windows~\cite{zhang2020syncwise}.
\end{enumerate}
\subsection{Advantages of the Proposed Top-down Approach}
A broad summary of the motivational study highlights the unrealistic nature of the assumptions surrounding the availability of granular micro-activity annotations made in the bottom-up approaches. Clearly, a more judicious approach to characterizing complex ADLs would be to \textbf{exploit the already available less ambiguous coarse-grain annotations that are far easier to obtain both in terms of time and cost}. However, developing such a top-down approach is not straightforward due to the lack of prior information regarding the temporal composition of different complex ADLs. Interestingly, while performing the motivational study, we also observed that most of these micro-activities produce completely different locomotive signatures, which, if correctly segmented and mapped in time, can be used to identify the micro-activities without external supervision. Motivated by this broad idea, we next develop \method{}, which utilizes unsupervised change-point detection followed by generalized zero-shot learning to identify the micro-activities that constitute a given macro-activity.
\section{Developing \method}
\label{overview}
\begin{figure*}[]
	\centering
	\captionsetup[subfigure]{}
	\begin{center}
		\subfloat[\label{fig:aquarezia}]{
			\includegraphics[width=0.55\columnwidth,keepaspectratio]{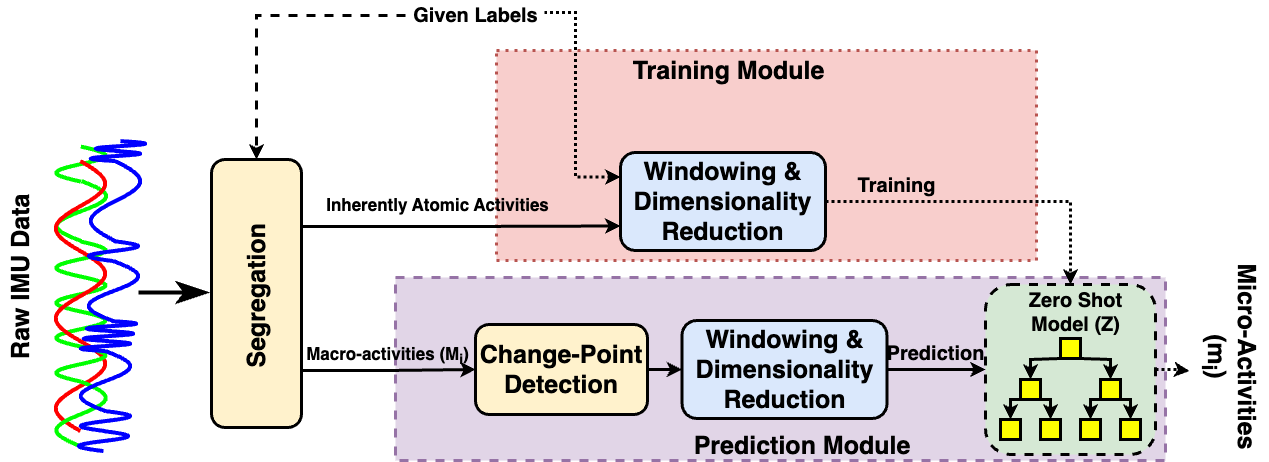}
		}
		\subfloat[\label{fig:zero_shot}]{
			\includegraphics[width=0.45\columnwidth,keepaspectratio]{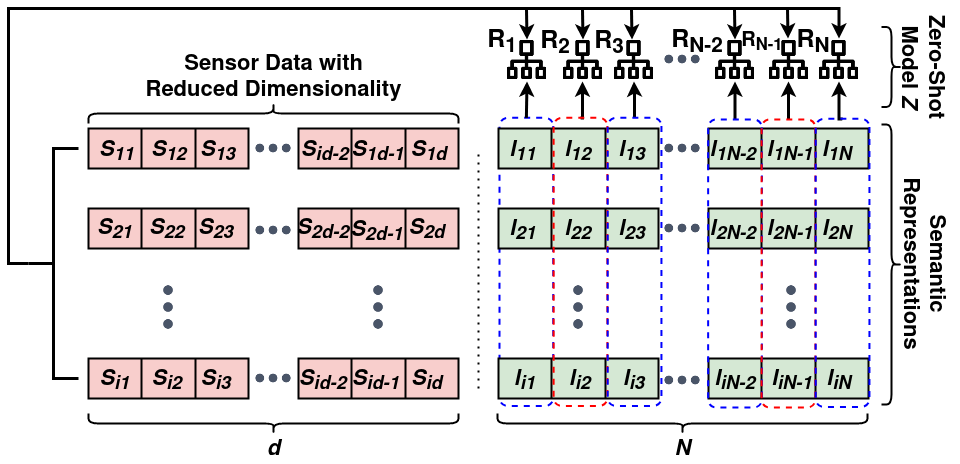}
		}
	\end{center}
	\caption{(a) High-level view of \method; (b) The zero-shot model $\mathbb{Z}$ during training phase. Here, $d$ and $\mathbb{N}$ represent the dimension of sensor data after dimensionality reduction and the dimension of the semantic representations, respectively.}
	\label{fig:method2}
\end{figure*}

Before continuing with the details of \method, we first briefly describe the overview of the proposed framework. As shown in Figure~\ref{fig:aquarezia}, the development of \method~involves three major phases -- (a) \textbf{Segregation:} In this step, we segregate the short-duration macro-activities based on a time-threshold $T$. This segregation is done based on the assumption that if any activity takes a sufficiently long time to complete, it has a high chance of containing multiple micro-activities. Following the segregation step, \method{} is composed of two major modules -- (b) \textbf{Training Module}: The responsibility of this module is to train the \method{} using zero-shot learning approaches after carefully capturing the sensor variability. (c) \textbf{Prediction Module:} Finally, followed by the training of \method{}, the next step is to utilize the model to identify the micro-activities from the change-point segmented IMU data for the macro-activities with duration $\ge T$. A detailed explanation of the training and the prediction modules follows.
\section{Training Module of \method}
\label{train}
We first start with the training phase of \method; the detailed steps are explained as follows.
\subsection{Creating the Training Set}
As shown in \figurename~\ref{fig:aquarezia}, the set of short-duration macro-activities, which have a duration less than $T$, constitute the training set $\mathbb{V}$. Essentially, the training set $\mathbb{V} = \{(u, a): u\in \mathcal{I}, a \in \mathcal{A}\}$ is a set of tuples $(u, a)$, where $\mathcal{I}$ is the set of input IMU data (here, accelerometer only) and $\mathcal{A}$ is the set of short-duration macro-activity labels.
Notably, for multi-sensor environments, where similar units of the same sensor (for example, locomotive sensors in this case) are attached to different body parts, it is absolutely critical to perform sensor selection to infer the activities correctly~\cite{zhang2013senstrack,keally2011pbn}. To resolve this concern, we, in the next step, employ a label-wise dimensionality reduction approach to reduce the input feature size while preserving the variations in the sensor data. The details follow.
\subsection{Windowing and Dimensionality Reduction}
Effective sensor selection often applies various supervised approaches to choose the proper sensors, which can improve the accuracy of the end model~\cite{zhang2013senstrack,keally2011pbn}. However, for approaches involving zero-shot modeling where label information may not be available apriori, completely supervised approaches may not be suitable. Following this understanding \method~employs  Uniform Manifold Approximation and Projection (UMAP)~\cite{mcinnes2018umap-software,SMG2020} so that during prediction, no supervised sensor selection approach is required to choose the sensors explicitly.

In a practical sense, this dimensionality reduction approach allows us to maximize the variance of the projected sensor data, potentially allowing us to focus more on the sensor units that vary more. For example, we seldom use different body parts other than our hands in a typical kitchen environment. Therefore, the sensor units attached to the hands are expected to vary more than the other IMU units attached to body parts like legs and torso. Interestingly, this technique also allows us to tackle problems like \say{change of hand}, where the subject shifts the task from one hand to the other in the same time interval.

However, in the training phase, a straightforward dimensionality reduction of the entire IMU data $\mathcal{I}$ may capture the variability present in the entire data, confounding all the variations captured by the sensor data for different activities. Understanding this, we perform a label-wise dimensionality reduction during the training phase. For this, we start by computing the magnitude of each IMU unit averaged across the time window of $1$ second. Once this is done, we perform dimensionality reduction on this window-averaged magnitude data separately for each short-duration macro-activity label. Formally, this transformed sensor data can be represented by $\mathbb{S}_i = \{s_{i_1},s_{i_2},s_{i_3},\hdots,s_{i_d}\}$, where $s_{i_j}$ is the transformed accelerometer data with dimensions reduced to $d$ (we fix $d=2$ in this paper). Naturally, the training data now takes the form $\hat{\mathbb{V}} = (\mathbb{S},\mathcal{A})$ which is next used to train the zero-shot model.
\subsection{Training the Zero-Shot Model}
Once the training dataset is ready, the next task is to train a model to predict the micro-activities, $\mathbb{M}^{t}_{t+\tau}$ for duration $[t,t+\tau]$. Notably, the micro-activities can be any activity that may or may not be seen by the model in $\mathbb{V}$, which definitely restricts the applicability of the standard supervised algorithms in this case. Understanding this, \method~employs a generalized zero-shot approach~\cite{larochelle2008zero} for modeling the task of predicting micro-activities. Typically, the zero-shot approach predicts the unseen labels using the information from the available labels in the training set. However, the existing conventional approaches~\cite{al2020zero,matsuki2019characterizing} of using zero-shot learning for activity predictions perform highly inaccurately~\cite{tong2021zero}.

\begin{table*}[]
	\centering
	\scriptsize
	\caption{A few example verb attributes defined in~\cite{Zellers2017ZeroShotAR}}
	\label{tbl:verb_attr}
	\begin{tabular}{|c|c|c|l|} 
\hline
\begin{tabular}[c]{@{}c@{}}\textbf{Verb}\\\textbf{Attributes}\end{tabular} & \begin{tabular}[c]{@{}c@{}}\textbf{Dimension of}\\\textbf{the Attribute}\end{tabular} & \textbf{Is Binary?} & \multicolumn{1}{c|}{\textbf{Description}}                                                                                                                                                                                                           \\ 
\hline
\begin{tabular}[c]{@{}c@{}}Aspectual\\Classes\end{tabular}                 & 1                                                                                     & N                   & \begin{tabular}[c]{@{}l@{}}Indicates whether a verb does not describe a change (like have),\\whether can be completed in short time (like open) or long time\\(like climb) or does not have a sense of completion at all (like walk).\end{tabular}  \\ 
\hline
\begin{tabular}[c]{@{}c@{}}Motion\\Dynamics\end{tabular}                   & 1                                                                                     & N                   & \begin{tabular}[c]{@{}l@{}}Energy level associated with the motion dynamics. Typical values\\can be high, low, medium.\end{tabular}                                                                                                                 \\ 
\hline
\begin{tabular}[c]{@{}c@{}}Body\\Involvements\end{tabular}                 & 5                                                                                     & Y                   & \begin{tabular}[c]{@{}l@{}}Movement of prominent body parts like head, hand, legs, torso,\\and other body parts.\end{tabular}                                                                                                                       \\
\hline
\end{tabular}
\end{table*}
\begin{table}
\scriptsize
\centering
\caption{Example verb attribute embedding for the activity \say{spray-pan}.}
\label{tbl:example_attr_embedding}
\begin{tblr}{
  row{even} = {c},
  row{1} = {c},
  row{3} = {c},
  cell{1}{1} = {c=3,r=2}{},
  cell{1}{4} = {r=2}{},
  cell{1}{5} = {r=2}{},
  cell{1}{6} = {r=2}{},
  cell{1}{7} = {r=2}{},
  cell{1}{8} = {c=5}{},
  cell{4}{1} = {c=12}{},
  vlines,
  hline{1,3-6} = {-}{},
  hline{2} = {8-12}{},
}
\textbf{Transitivity}         &   &   & \textbf{Aspect} & \textbf{Motion} & \textbf{Time} & \textbf{Social} & \textbf{Bodyparts } &               &               &                &                 \\
                              &   &   &                 &                 &               &                 & \textbf{Arms}       & \textbf{Head} & \textbf{Legs} & \textbf{Torso} & \textbf{Other } \\
1                             & 1 & 1 & 3               & 3               & 1             & 0               & 1                   & 0             & 0             & 0              & 0               \\
\textbf{Effect on Arguments } &   &   &                 &                 &               &                 &                     &               &               &                &                 \\
0                             & 1 & 0 & 0               & 0               & 1             & 1               & 0                   & 0             & 1             & 1              & 0               
\end{tblr}
\end{table}
We, in this paper, address this problem by choosing different types of \emph{semantic representations} of the activities as the latent embeddings for training the model. Notably, we broadly compare (a) motion-based semantic attributes defined in~\cite{lara} and (b) a more generalized verb-attribute embedding defined in~\cite{Zellers2017ZeroShotAR}. Unlike the motion-based semantic attributes (See Section~\ref{lara_dataset}), the verb attributes are generated from the existing dictionary definitions (some example verb attributes shown in~\tablename~\ref{tbl:verb_attr}). Additionally, the generalized verb attributes also add verb templates while defining the verb attributes. As defined in~\cite{Zellers2017ZeroShotAR}, this addition of templates allows us to consider the context of a word. This, in turn, enables us to separate similar labels with dissimilar actions in label space. For example, the verb \say{open} can be used with \textit{door} as well as a \textit{packet}, albeit the actions are entirely different. Verb templates enable us to keep these labels separate by introducing the context as a verb template for adequately defining the actions (an example verb attribute embedding for the activity label \say{spray-pan} is shown in \tablename~\ref{tbl:example_attr_embedding}).

Subsequently, with the usage of these semantic attributes, any $i$\textsuperscript{th} activity label from $\mathcal{A}$ can be represented as $\mathcal{A}_i = $ semantic attributes of the short-duration macro-activity label $a_i = \{l_{i_1}, l_{i_2}, \hdots, l_{i_\mathbb{N}}\}$
, where $\mathbb{N}$ is the dimension of the semantic attribute embeddings. Considering these semantic representations as the labels, the final zero-shot model $\mathbb{Z}$ can be defined as $\mathbb{Z} = \{$Model $\mathcal{R}_j| \forall j$ $\in$ $\{1,2,\hdots,\mathbb{N}\}\}$ where any model $\mathcal{R}_j$ is trained using the transformed sensor data $\mathbb{S}$ and the latent embedding $l_j$ from the semantic representation for the short-duration macro-activities in $\mathcal{A}$ (as shown in \figurename~\ref{fig:zero_shot}).
\section{Prediction of the Micro-Activities}
\label{pred}
Once the model is trained, the next task is to identify the micro-activities $\mathbb{M}^{t}_{t+\tau}$ for time duration $[t,t+\tau]$. However, it is difficult to characterize and predict the different micro-activities separately without any prior information about the number of micro-activities present within a macro ADL. Thus, there is a need to segregate the entire duration into segments where each segment potentially corresponds to a separate micro-activity. To obtain these segments, we first use an unsupervised change-point detection scheme. Subsequently, we use the locomotive signatures corresponding to these individual segments to obtain the micro-activities from the trained zero-shot model $\mathbb{Z}$. The details follow.
\subsection{Change-Point Detection}
\label{change_pt}
Directly looking into the locomotive signatures for the entire duration of any complex activity may not provide the required information regarding the micro-activities. Moreover, there can be more than one micro-activities that may be present in that entire duration. Interestingly, IMU signatures are known to precisely capture the changes caused by the activities~\cite{chatterjee2020laso}. Following this understanding, we first segment the IMU data $\mathcal{I}^{t}_{t+\tau}$ corresponding to a macro-activity label $\mathbb{L}^{t}_{t+\tau}$ for the duration $[t,t+\tau]$ according to the change in activities using an unsupervised approach. For this, we rely on unsupervised change-point detection approaches like~\cite{rulsif,killick2012optimal,truong2020selective} for detecting changes in the IMU signatures. We provide a detailed comparison of different change-point detection algorithms in Section~\ref{eval}.
\subsection{Windowing and Dimensionality Reduction}
Once the change-points are obtained, we compute the average magnitude of the IMU signatures across the change-point windows. Next, we apply a dimensionality reduction technique similar to the training phase. As we do not have the exact micro-activity labels during the prediction phase, we use UMAP to reduce the dimension to $d$ for the duration $[t,t+\tau]$. Once this is done, the final step is to predict the micro-activities $\mathbb{M}^{t}_{t+\tau}$ for each time window $[t,t+\tau]$.
\subsection{Detecting Micro-Activities}
For the prediction, we query each trained model in $\mathbb{Z}$ with an instance of the transformed IMU data (obtained from the previous step) and obtain a set of predictions regarding the semantic representations. Subsequently, the micro-activities $\mathbb{M}^{t}_{t+\tau}$ can then be derived directly from the interpretation of each semantic in the embedding space. Alternatively, these micro-activities $\mathbb{M}^{t}_{t+\tau}$ can also be identified by observing the neighboring attributes in the embedding space. At the same time, their meaning can be extracted using the set of known attributes. For example, for the verb attributes, the meaning of the attribute can be obtained from the embedding of known verbs that are in proximity to the generated verb attribute.
\section{Obtaining Ground-truth Micro-activity Annotations}
\label{grnd_annotation}
For evaluating \method in a principled manner, we first re-annotate and extend the macro-activities with the \emph{ground-truth micro-activities}. The steps for the kitchen and the LARa datasets are discussed as follows.
\subsection{Kitchen Dataset}
As the original dataset only had recipe-level annotations for macro-activities, we next re-annotated the data with ground-truth micro-activity annotations. However, such a dedicated re-annotation to label micro-activities may lead to the generation of unnecessary micro-activity labels that may not be realistic to identify and label. Therefore, we carefully designed a survey-based setup where we could ensure the quality of the ground-truth micro-activity labels and filter out any spurious gesture or activity getting annotated. The details follow.
\subsubsection{Procedure and Setup}
\label{reannotate}
In order to develop the understanding in a principled manner, we first obtain the micro-activity labels with the help of a human annotator. We achieve this by re-annotating the macro-activities (where $T > 10$s) from the kitchen dataset, where we expand the existing macro-activity labels to the finer micro-activity labels. For this task of re-annotation, we instruct the independent annotator to expand the existing macro-labels and to provide labels with higher granularity. For example, the macro activity label \say{take baking pan} for the subject S01 is expanded and re-annotated with a higher level of granularity as \say{open bottom cupboard, take the pan and finally close the cupboard}, which essentially provides us a collection of various ground-truth micro-activities. Additionally, during this process, we ensure that all these ground-truth micro-activities are mapped to their respective time windows.
\subsubsection{Removing Spurious Ground-truth Micro-activity Labels}
A dedicated task of re-annotating the data with a higher granularity of labels can often lead to situations where some macro-activities are extended with spurious activity labels that may not be important for any activity recognition system to identify. To specifically identify and characterize such spurious annotations, we conducted a focused study with $19$ participants ($4$ females and $15$ males). All these participants are computer science researchers or faculty members in established universities across the globe. We specifically choose such a set so that the participants understand the challenges in providing annotations and mere linguistic variations\footnote{Notably, all the participants involved in the study are non-native English speakers.} do not impact their judgment while demarcating a specific annotation as informative. Summarily, the study was conducted to answer the following research questions:
\begin{enumerate}
	\item \textit{Does higher granularity always improve annotation's quality?} 
	\item \textit{Is this micro-activity annotation necessary to be identified?}
\end{enumerate}
To answer these questions, we design the study as a questionnaire with visual inputs taken from the activities for $5$ subjects from the kitchen~\cite{cmummac} dataset ($2$ subjects with the highest and $3$ subjects with the lowest number of activities with a duration $>10$ seconds). The visual inputs are taken from the ground-truth first-person videos in the dataset. The specific portions where those activities are being performed have been clipped. Those clips are then attached to the form along with two labels in human-readable format, one given in the dataset (marked as \textbf{A}) and another with a re-annotated detailed label (marked as \textbf{B}). For example, the clip where the user S04 sprays the baking pan in the dataset is followed by two labels -- \textbf{(A)} \say{spray-pan} (the default label from the dataset) and \textbf{(B)} \say{open cap, shake the spray can then spray the pan with it, and finally close the cap} (the re-annotated label). Finally, for each such clip with two labels, we ask the participants the following questions --
\begin{enumerate}
	\item \textit{Which of the two given labels more appropriately describes the activity performed in the video clip?}
	\item \textit{How do you justify accepting or rejecting the re-annotated label?}
\end{enumerate}
The first question is a simple multiple-choice answering type with options for selecting or rejecting the detailed re-annotated label. Additionally, the participants were given choices to judge whether the given labels were equally informative considering the activity shown in the clip. Following this, the second question focused on why the participants feel that the re-annotation should (or should not) be done. This is a subjective question, and the participants were asked to provide specific reasons that justify their choice of labels. This entire questionnaire is followed for all the $5$ chosen subjects and at least $5$ independent participants\footnote{Throughout the paper, we use the term \textit{subject} to indicate the users who performed the activities in the respective datasets, \textit{annotators} to indicate the peoples who have annotated the data with some label, and \textit{participants} to indicate the contributors who have participated in this survey done by us.} answer each questionnaire to achieve consensus.

Through this survey, we obtained a series of interesting observations. Some key insights that allow us to filter spurious ground-truth micro-activity annotations are as follows.
\begin{enumerate}
    \item \textbf{Micro-activity annotations that add no additional information:} For a very few cases , the participants have also felt that the detailed label carries the same information as the coarse-grain label, and it might not be necessary to re-annotate in such cases. These cases, include activities like \say{pour big bowl into baking} (original label from the kitchen dataset) where the detailed re-annotated label specified was \say{pour big bowl into baking while moving it}. Clearly, the detailed label does not add much to the explanation provided by the existing label.
    
    \item \textbf{Activity induced body movements and impact of multitasking:}Multitasking of activities is already known to confuse the annotators~\cite{cmummac}. This often leads to the annotators designating one of the multiple tasks as primary and subsequently demarcating that task as the final activity label. For example, participant P12 felt for the activity \say{pour big bowl into baking pan} with a re-annotated label as \say{pour big bowl into baking pan while moving it}.\\

    \textit{B is somehow obvious; to pour the content, typically, a person will try to spread the contents equally across the baking pan, so will move it.}\\
    
    A similar point is also made by P15, who thinks that certain gestures and movements are inherently present and are induced by the overall activity.\\
    
    \textit{"Shaking" can be considered as an activity-induced phenomenon. Consider pouring the brownie holding the brownie packet completely upside down instead of in a slanted position. Packet shaking, wiggling, and even squeezing can be considered as noise.}\\

    However, the participants did not find all such instances where the subjects were multitasking as unnecessary to be re-annotated. Clearly, there were situations where certain activities done in parallel were deemed meaningful and the participants felt it was required that these are included in re-annotation. For example, \say{pouring the contents}, and at the same time, \say{mixing the contents of the bowl}. In such cases, participants like P5, P11, P13, P14, and P18 feel that the detailed label \textbf{B} which includes the action of \say{pouring the contents} is more relevant. In particular, P14 expresses his perspective by explaining as follows.\\

    \textit{Clearly, there are two actions happening simultaneously - stirring and pouring the contents of a bag, which only label B explains.}
\end{enumerate}

Finally, these spurious micro-activity labels are removed, and the mapped meaningful micro-activity labels are used for evaluating \method{} on the kitchen dataset.
\subsection{LARa Dataset}
For the LARa dataset, we use the already timestamped 18 semantic attributes as the ground-truth micro activity annotations (Section~\ref{lara_dataset}). Notably, for the LARa dataset, we set the time threshold $T>15$s for further investigation regarding the micro-activities. 
\section{Evaluation Metrics and Setup}
\label{setup}
Next, we define the metrics that we implement to evaluate the different components of \method. The details follow.

\subsection{Accuracy of Micro-activity Identification}
In order to evaluate the accuracy of the zero-shot model, we compare the predicted micro-activities with the ground-truth micro-activity labels using the following two standard metrics.

1. \textbf{Quantitative Evaluation:} We compute the micro F\textsubscript{1}-score of the attribute vector of the predicted micro-activities and the latent embeddings of the ground-truth micro-activities.
	
2. \textbf{Subjective Evaluation:} Considering $m_i \in \mathbb{M}^{t}_{t+\tau}$ as the micro-activity predicted by \method, we identify the top-5 verbs (obtained from the verb corpus~\cite{Zellers2017ZeroShotAR}), which are closest in the embedding space with $m_i$. This provides a subjective evaluation of the predicted micro-activity $m_i$ in realistic scenarios.

\subsection{Accuracy of Change-Point Detection:}
For evaluating the performance of the change-point detection technique, we define two metrics as follows. define the metrics for evaluating \method{}, and finally discuss the implementation details of \method{}.

1. \textbf{Annotation Error} (AE)~\cite{truong2020selective} measures the absolute difference in the number of change-points observed in the ground-truth and the estimated number of changes computed by \method{} (see Section~\ref{change_pt}). Essentially, a low annotation error will allow \method{} to correctly extract the activity patterns in the individual segments, which would allow the zero-shot model to make more accurate predictions.

2. \textbf{Mean Absolute Error} (MAE)~\cite{deldari2020espresso} measures the difference between the ground-truth segment time and the nearest estimated segment time (in secs) identified by \method. At least MAE ensures that the segments found by the change-point detection module are close to the actual boundaries observed in the ground truth, which in turn reduces the temporal error locating the micro-activity within the macro-activity.

\subsection{Implementation and Hyperparameters}
For the individual models $\mathcal{R}_j$ in $\mathbb{Z}$, we choose Random Forest Classifiers (with the number of estimators = $100$ and maximum depth = $34$) for the kitchen dataset. However, as the motion-based embeddings for the LARa dataset are binary, unlike the generalized verb attribute embeddings, we design each $\mathcal{R}_j$ as Support Vector Machines (with radial basis function kernel). For the unsupervised dimensionality reduction we use the implementation of UMAP available in \texttt{umap-learn}\footnote{\url{https://github.com/lmcinnes/umap}} with number of neighbors = $2$ during the prediction phase, however, for the training phase we keep it at the default value of $15$.
\subsection{Baseline Change-Point Detection Schemes}
To compare the performance of the change-point detection approach we consider two other well-known unsupervised change-point detection algorithms as baselines. The details of their implementation is as follows.

1. \textbf{Relative Unconstrained Least-Squares Fitting (RuLSIF)}~\cite{rulsif}: RuLSIF compares the $\alpha$-relative Pearson Divergence Estimate across the windows of the sensor signals. For implementing the same we use the standard \texttt{densratio}\footnote{\url{https://github.com/hoxo-m/densratio_py}} package and set $\alpha=0.01$. However, RuLSIF only generates change-point scores which are difficult to interpret directly. To resolve this, we simply perform an unsupervised clustering of the scores for demarcating the actual changes~\cite{chatterjee2020laso}.

2. \textbf{Pruned Exact Linear Time (PELT)}~\cite{truong2020selective} The core of this algorithm relies on finding the optimal number of segments, fine-tuned by the hyperparameter~\emph{penalty}, in the signal without any prior information. Additionally, the algorithm has a linear running time which gives a considerable speedup. Notably, the accuracy of this algorithm is heavily constrained by the hyperparameter penalty, whenever the algorithm commits an error or detects a spurious change-point. In essence, this parameter balances out the goodness-of-fit, and subsequently, a high penalty value may force this algorithm to identify only significant changes~\cite{truong2020selective}. we use the implementation available in \texttt{ruptures}\footnote{\url{https://github.com/deepcharles/ruptures}} package with a penalty of $100$ and $50$ for the kitchen and the LARa datasets, respectively.

3. \textbf{Kernel Change-Point Detection (KernelCPD)}~\cite{celisse2017new}: We implement the KernelCPD using the \texttt{ruptures} package. For this algorithm we choose a linear kernel. Similar to PELT, this algorithm also allows fine-tuning with penalty. Here as well, we set the value of penalty as $100$ and $50$ for the kitchen and LARa datasets, respectively.

\section{Performance Evaluation of \method}
\label{eval}
In this section, we evaluate the performance of \method~by measuring the accuracy of micro-activity prediction. Furthermore, we also investigate the performance of \method{} by looking into the accuracy of the change-point detection, which serves as a core module behind observing the activity boundaries. Finally, we also perform a qualitative analysis using \method{} with SOTA LLMs to obtain more explainable micro-activity labels from the semantic embeddings.
\subsection{Accuracy of Micro-activity Identification}
\begin{figure*}
	\centering
	\begin{center}
		\subfloat[\label{fig:accuracy_pca}]{
				\includegraphics[width=0.45\columnwidth,keepaspectratio]{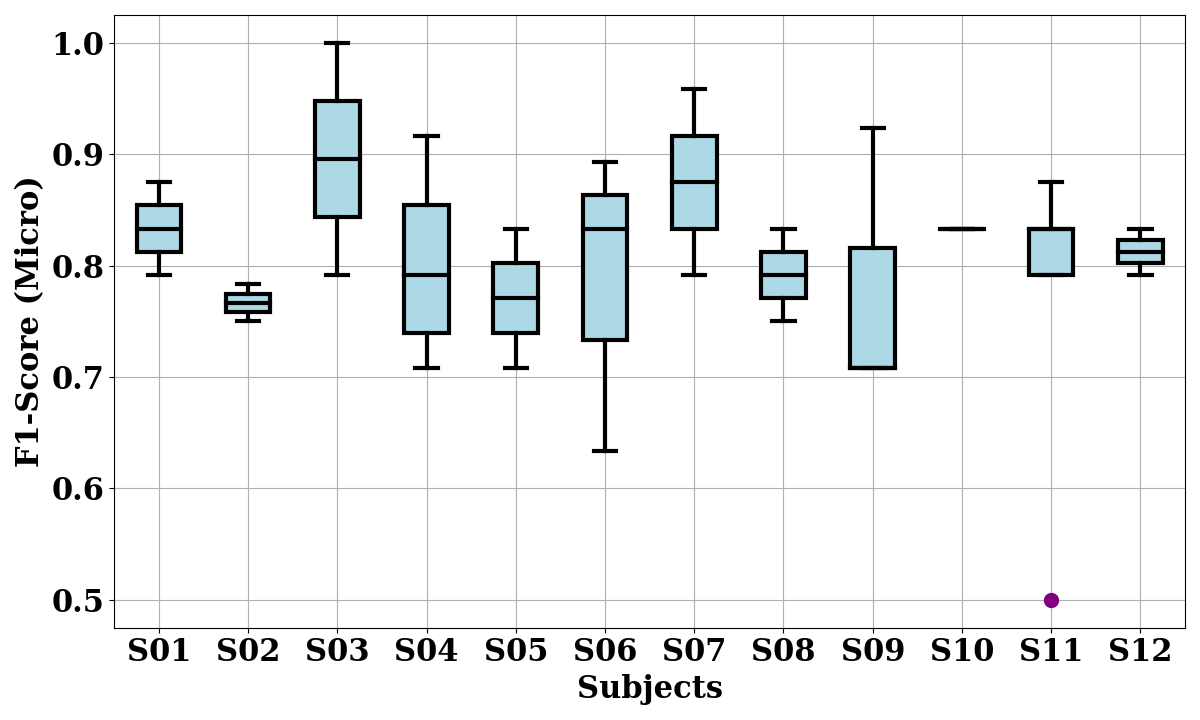}
		}	
    \subfloat[\label{fig:example1}]{
			\includegraphics[width=0.55\columnwidth,keepaspectratio]{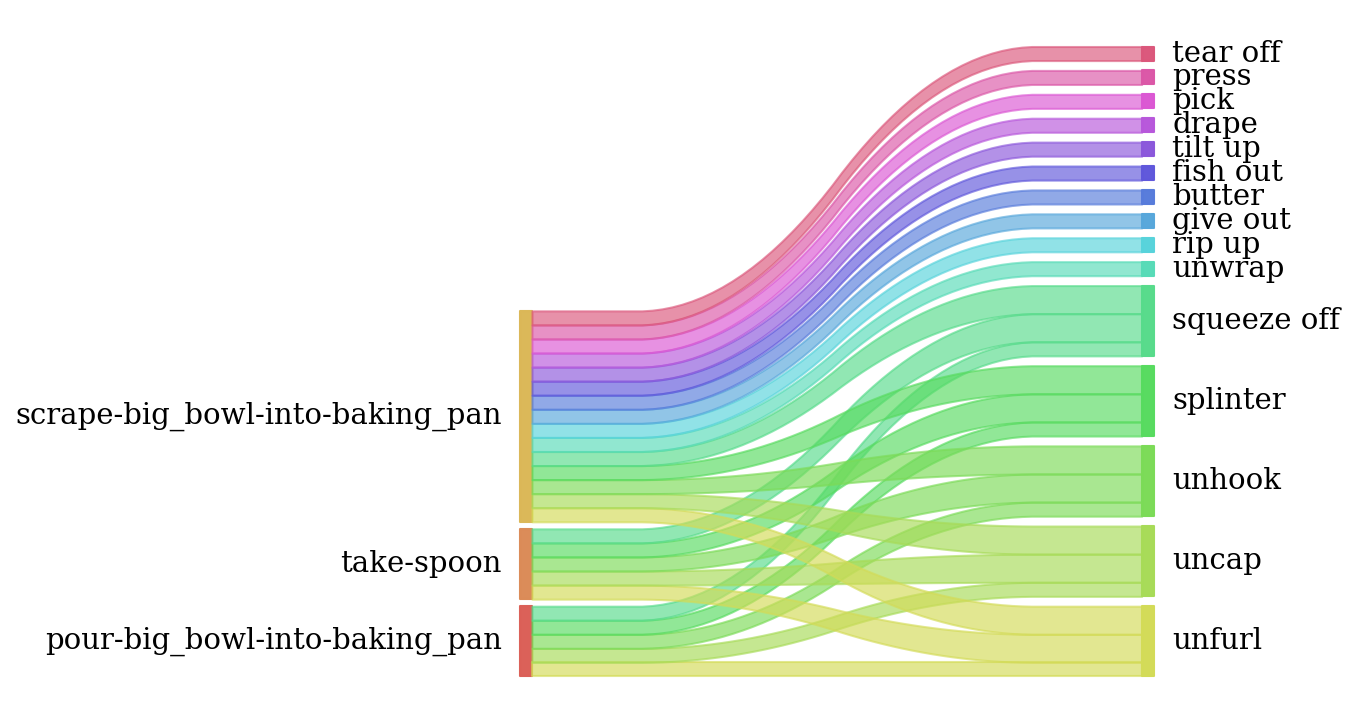}
		}
  
  	\subfloat[\label{fig:example2}]{
			\includegraphics[width=0.63\columnwidth,keepaspectratio]{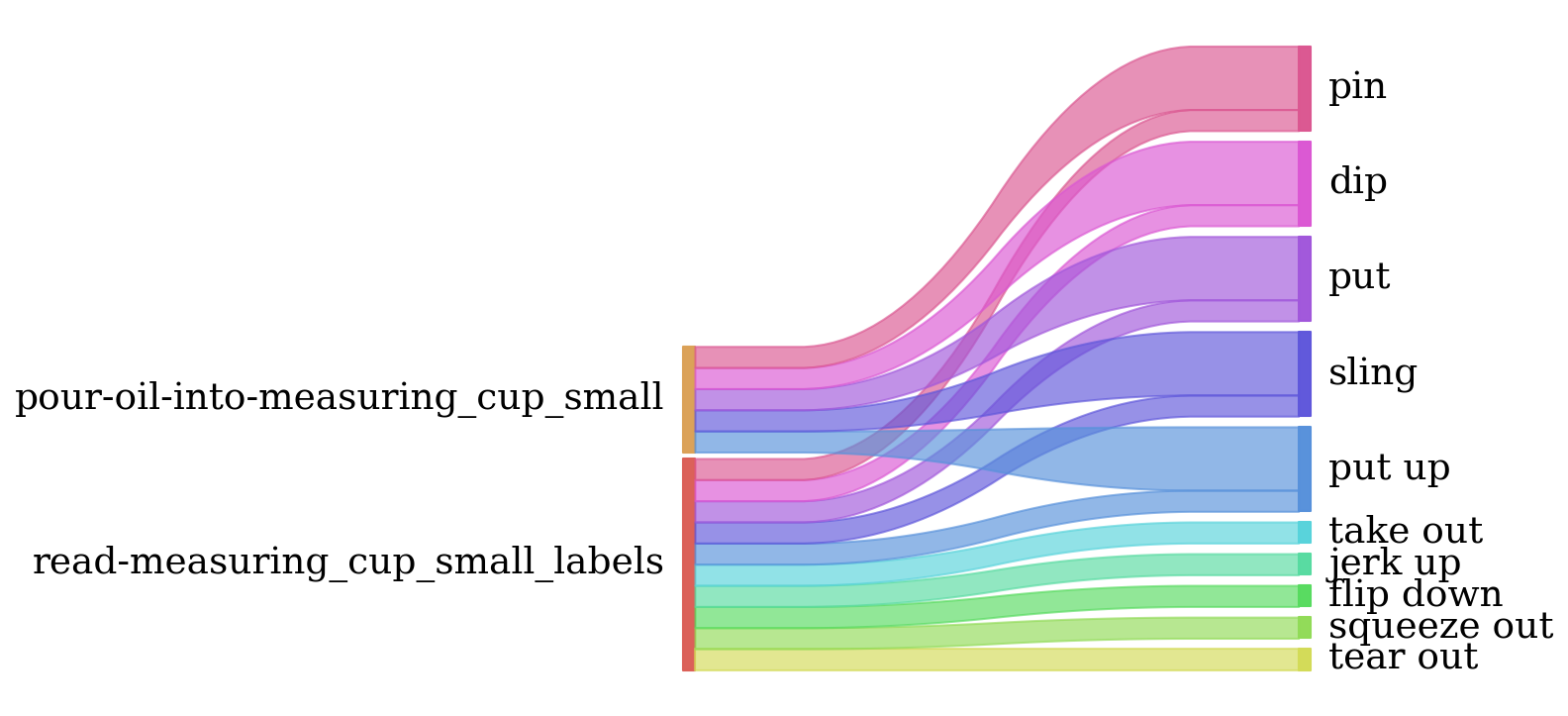}
		}
	\end{center}
	\caption{(a) F\textsubscript{1}-score of micro-activity prediction for the kitchen dataset and Top-5 closest verbs (right) in embedding space for the ground-truth micro-activities (left) (b) S06: \say{pour big bowl into baking pan} and (c) S02: \say{pour oil into measuring cup small}}
	\label{fig:accuracy}
\end{figure*}

\textbf{Kitchen Dataset:} In~\figurename~\ref{fig:accuracy_pca}, we observe that for most of the subjects, the zero-shot model $\mathbb{Z}$ predicts the micro-activities (in terms of verb attributes) with an appreciable median F\textsubscript{1}-score of $>0.75$, for most of the subjects. Notably, the body movement-based verb attributes are predicted with a F\textsubscript{1}-score of $1$ accuracy for all cases, except for one corresponding to subject S02, where the mean body movement F\textsubscript{1}-score drops to $0.73$ $(\pm 0.19)$. A deeper investigation of this case reveals that the change-point detection scheme first fails to identify the ground-truth micro-activity of \say{reading the measuring cup labels}, which degrades the performance of the zero-shot model as well.
Interestingly, the existing annotations obtained directly from the kitchen dataset~\cite{cmummac} also overlooked this intermediate micro-activity \say{reading} the labels, albeit the activity \say{reading} has been one of the recognized activities in their predefined corpus.
Nevertheless, we conclude from the quantitative evaluation that \method~can predict micro-activities appreciably.

In the subjective evaluation, we next examine the potential of the zero-shot model in correctly identifying the relevant micro-activities performed in the kitchen environment. For example, in \figurename~\ref{fig:example1}, for the macro-activity \say{pour big bowl into baking pan} (for subject S06) \method~correctly predicted the micro-activities like \say{butter} which resembles the action for the ground-truth micro-activity \say{scraping of the big bowl into the baking pan} followed by the predicted micro-activities like \say{squeeze off} which approximately resembles the action for the ground truth micro-activity \say{pour}.
\begin{figure*}
	\centering
	\begin{center}
		\subfloat[\label{fig:accuracy_lara}]{
				\includegraphics[width=0.48\columnwidth,keepaspectratio]{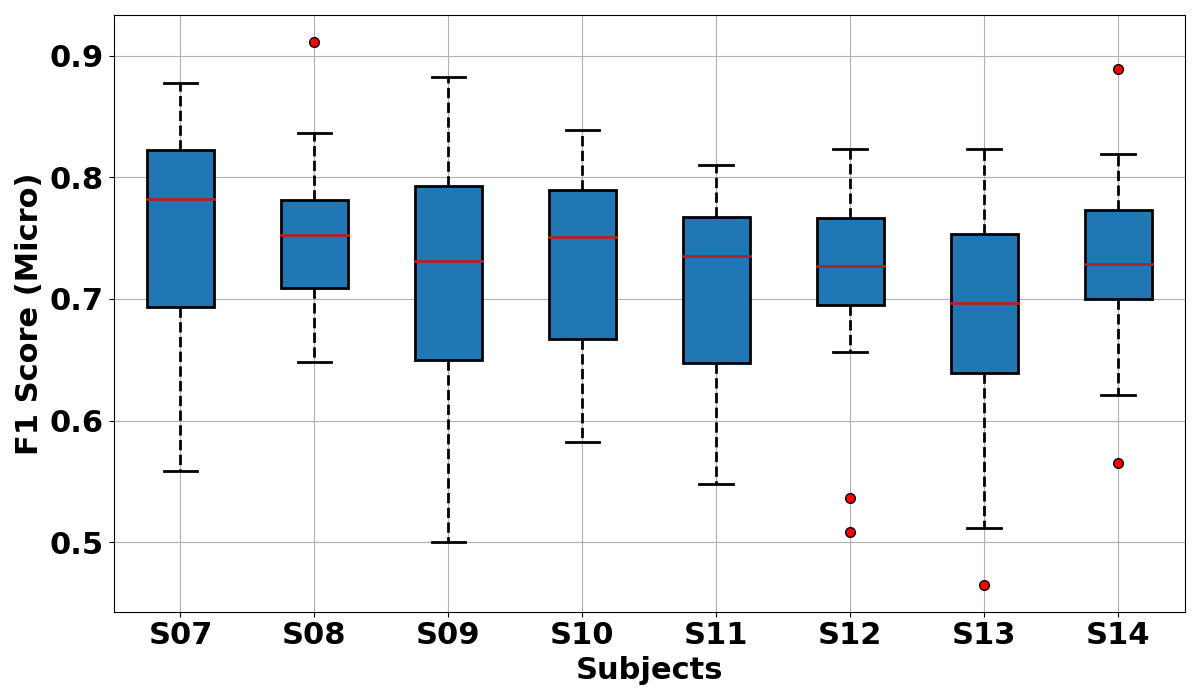}
		}	
  	\subfloat[\label{fig:accuracy_lara_movement}]{
			\includegraphics[width=0.48\columnwidth,keepaspectratio]{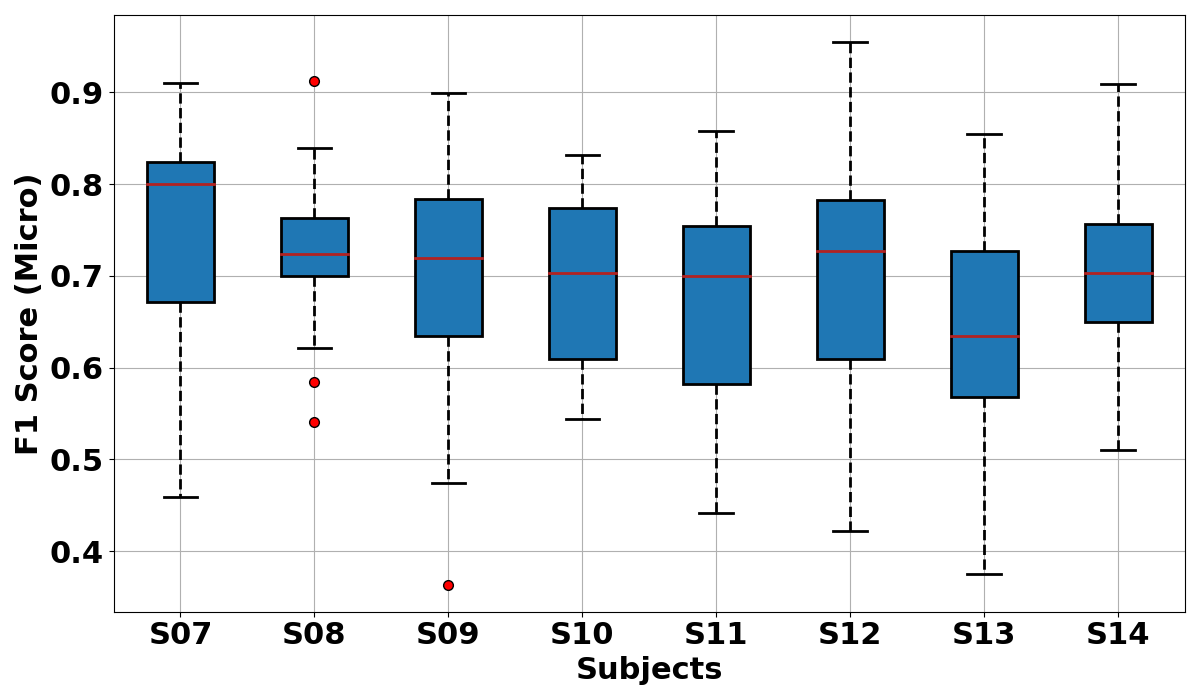}
		}
	\end{center}
	\caption{F\textsubscript{1}-score of predicting the motion-based semantic embeddings of the micro-activities across (a) all embeddings (b) only body movement based embeddings}
	\label{fig:accuracy_lara}
\end{figure*}

However, a deeper investigation reveals that completely irrelevant micro-activities may also appear in the top 5 predictions in some cases. For instance, in \figurename~\ref{fig:example2}, for the ground-truth micro-activity \say{pour oil into measuring cup small} (for subject S03), we observe that many irrelevant verbs (say, pin, sling) appear in the top-5 predictions. This is primarily because of these irrelevant yet ``universal'' neighboring verbs in the embedding space. This limitation is also known as hubness, which is already known to plague zero-shot models~\cite{al2020zero}. Importantly, we still note that the relevant micro-activities, in this case, \say{squeeze out}, and \say{flip down} are also predicted, demonstrating the elegance of \method.

\textbf{LARa Dataset:} Like the kitchen dataset, we analyze the performance of \method{} in predicting the motion-based semantic attributes defined in the LARa dataset. From \figurename~\ref{fig:accuracy_lara}, we observe that here as well \method{} can predict the exact motion-based embeddings with a median F\textsubscript{1}-score $\approx 0.75$ across all the subjects. Notably, these motion-based embeddings, as shown in Section~\ref{lara_dataset}, contain more sophisticated body movements and also record certain embeddings like \say{Bulky Unit}, \say{Handy Unit}, and \say{Computer} which are much difficult to predict considering the inputs from inertial sensors. Nonetheless, from both \figurename~\ref{fig:accuracy_lara} and \figurename~\ref{fig:accuracy_lara_movement}, we conclude that \method{} can capture sophisticated body movements and detect micro activities using with appreciable accuracy.
\subsection{Accuracy of Change-Point Detection}
\label{eval_change_point}
\begin{figure*}
	\centering
	\subfloat[Kitchen -- AE\label{fig:adnc1}]{
		\includegraphics[width=0.48\columnwidth, keepaspectratio]{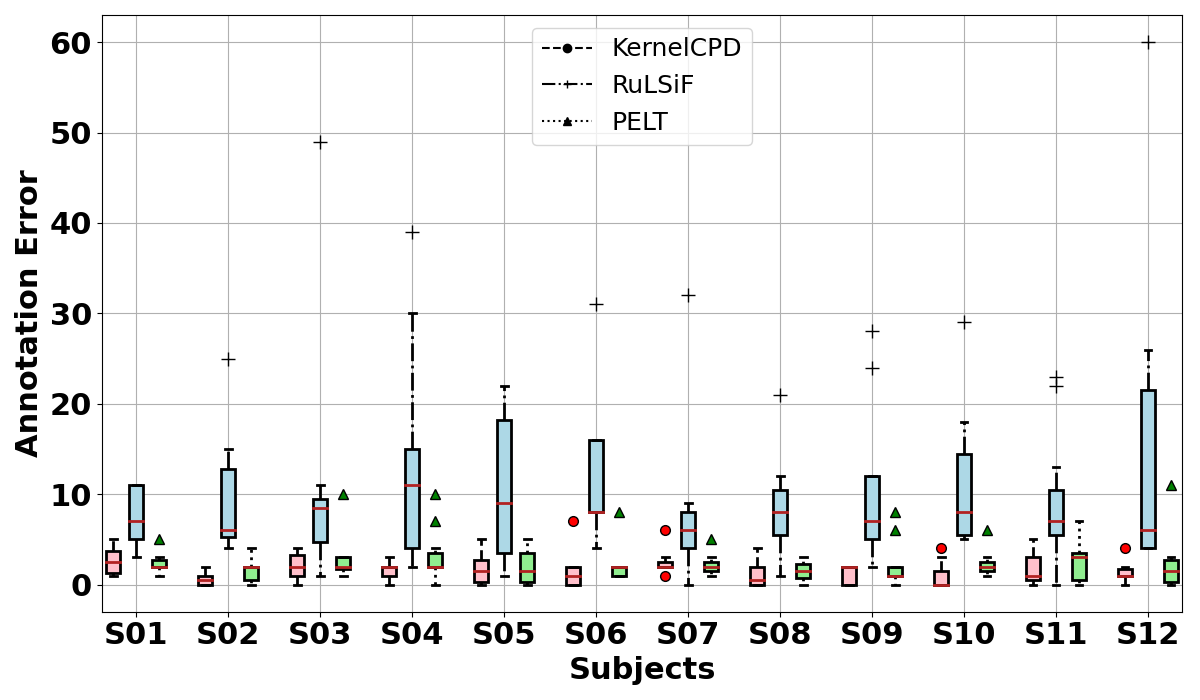}
	}
	\subfloat[Kitchen -- MAE\label{fig:mae1}]{
		\includegraphics[width=0.48\columnwidth, keepaspectratio]{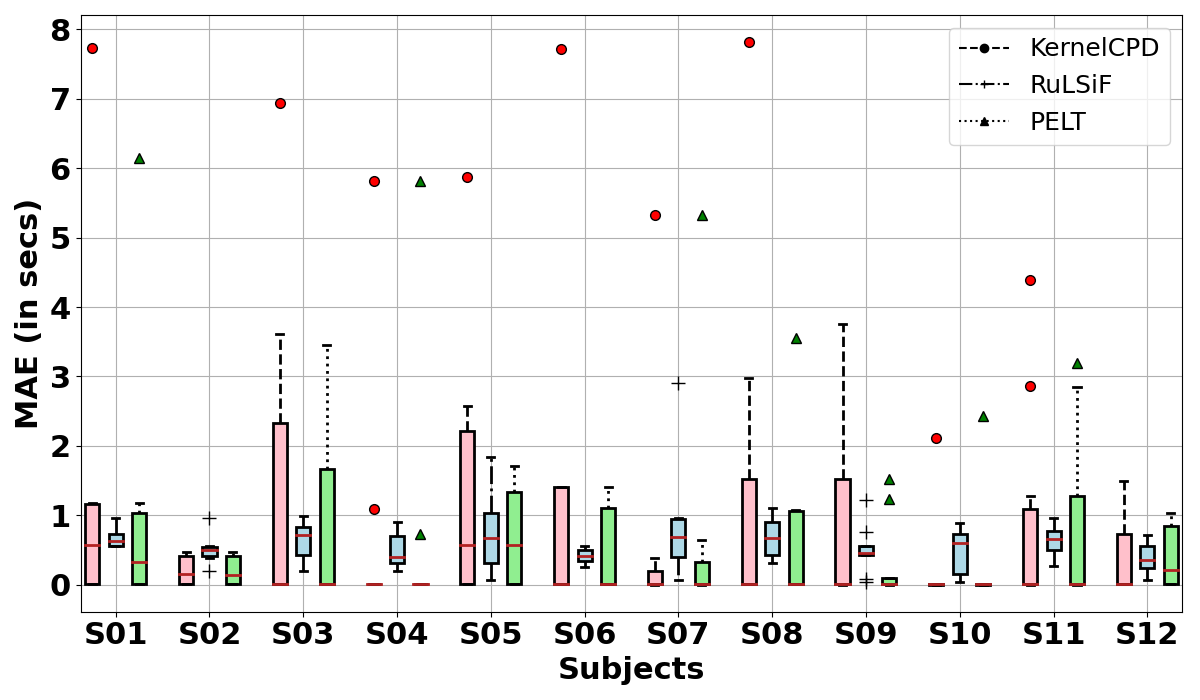}
	}	
 
	\subfloat[LaRa -- AE\label{fig:adnc2}]{
		\includegraphics[width=0.48\columnwidth, keepaspectratio]{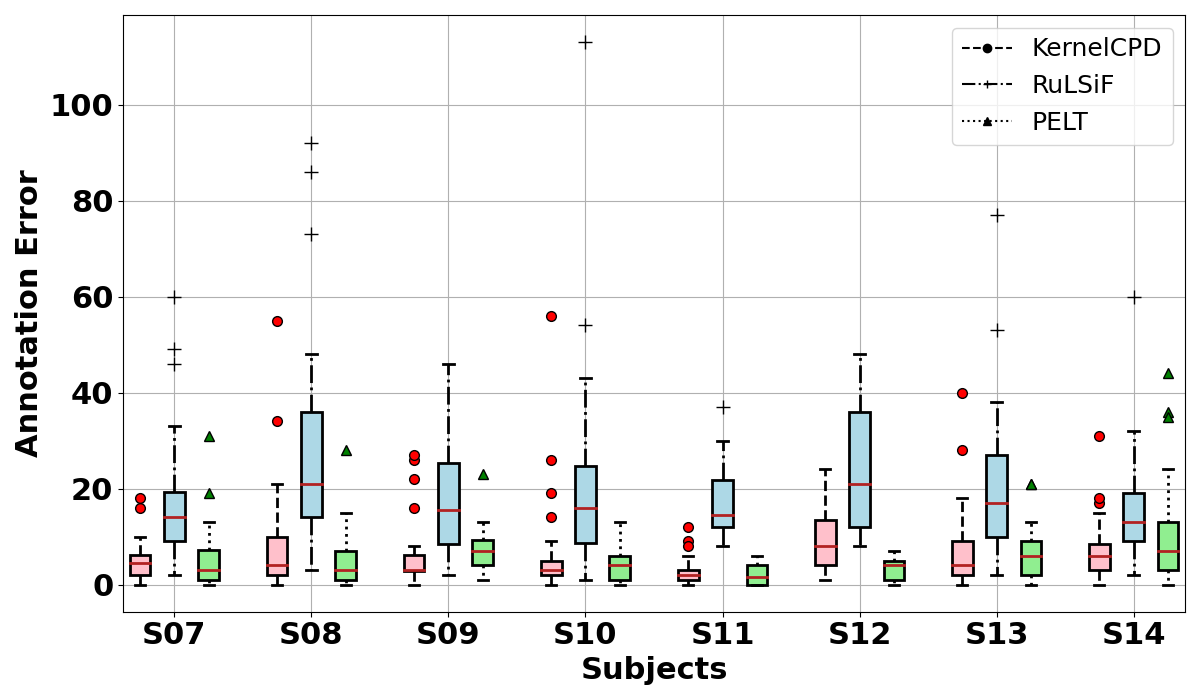}
	}
	\subfloat[LaRa -- MAE\label{fig:mae2}]{
		\includegraphics[width=0.48\columnwidth, keepaspectratio]{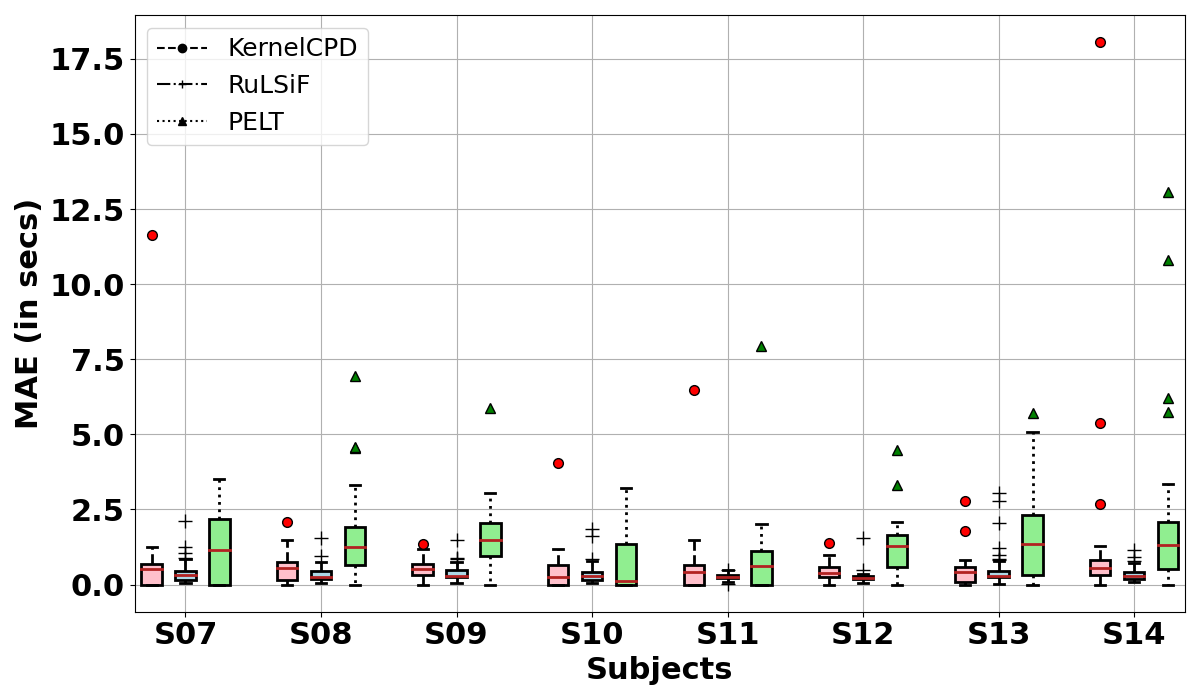}
	}
	\caption{Evaluation of the change-point approaches in terms of AE and MAE for (a), (b) - Kitchen and (c), (d) - LARa Datasets}
	\label{fig:change_pt_eval}
\end{figure*}
We first start by investigating the impact of change-point detection using AE and MAE as the metrics. In~\figurename~\ref{fig:change_pt_eval}, we observe that PELT performs optimally for both datasets by maintaining a low AE while keeping the MAE $<1$ second for most cases. Interestingly, we observe that KernelCPD and RuLSIF are somewhat data-dependent and perform poorly (in AE, MAE, or both) in at least one of the datasets. Nevertheless, it is comforting for us to see that the MAE mostly stays below $3$ seconds (regardless of the dataset and the change-point algorithm used), thus identifying the hidden micro-activities close to their exact ground-truth time windows.

\subsection{Opportunities of using \method{} with LLMs: A Qualitative Analysis with GPT-3.5}
One key advantage of using the \method{} is that it relies on attribute embeddings to identify the hidden micro-activities. As a result, \method{} could predict robust embeddings that capture various body movements and gestures, even for those activities that were not seen as a part of the training data. Additionally, with latent embeddings like verb attributes, \method{} can predict meaningful embeddings. This, in turn, allows us to use \method{} with SOTA LLMs like GPT-3.5~\cite{openai2023gpt4} to identify the micro-activities from the predicted verb attribute embeddings. To evaluate this possibility of using \method{} with LLMs, we first convert a set of predictions made by \method, in terms of verb attribute embedding, into queries\footnote{We do not use any form of sophisticated query processing or prompt tuning for this.}. For example, for a predicted verb-attribute embedding for the micro-activity like this \say{1,0,1,3,3,1,2,1,0,0,0,0,0,0,0,0,0,0,0,1,0,0,0,0,1,0,0,0,0,0} the generated query after adding the information about the encompassing macro-activity becomes:\\

\say{tell me a kitchen activity which is done while pouring the contents of a big bowl containing brownie mixture into a baking pan that is intransitive, can be used in the form of something, shows achievement, requires medium motion, requires time in order of seconds, is performed in solitary, which requires arms to be used for the action but not the head, legs, torso or any other body parts, which changes the external world and also the state of the object.}\\

\tablename~\ref{tbl:chatgpt} shows the summary of the response generated by GPT-3.5 for the aforementioned queries. The primary observations from this investigation show the potential of using frameworks like \method{} with the LLMs to extend the existing label space to allow more granular activity sequences, which in turn can help identify more complex ADLs.

\begin{table}[]
	\centering
	\scriptsize
 \caption{Using GPT-3.5 to identify the micro-activities from the generalized verb-attribute embeddings predicted by \method.}
 \label{tbl:chatgpt}
	\begin{tabular}{|l|l|l|l|}
		\hline
		\multicolumn{1}{|c|}{\textbf{\begin{tabular}[c]{@{}c@{}}Ground-truth\\ Macro-activity\end{tabular}}} &
		\multicolumn{1}{c|}{\textbf{Query}} &
		\multicolumn{1}{c|}{\textbf{\begin{tabular}[c]{@{}c@{}}Response\\ Micro-activity\end{tabular}}} &
		\multicolumn{1}{c|}{\textbf{\begin{tabular}[c]{@{}c@{}}Original\\ Observation\end{tabular}}} \\ \hline
		\multirow{4}{*}{stir big bowl} &
		\begin{tabular}[c]{@{}l@{}}tell me a kitchen activity related verb which is done while stirring a \\ big bowl containing brownie mixture which is not transitive can be used \\ in the form something, shows achievement, requires medium motion, \\ requires time in order of seconds, is performed in solitary, which \\ requires arms to be used for the action but not head, legs, torso or any\\ other body parts, which moves the object\end{tabular} &
		mix &
		\multirow{4}{*}{\begin{tabular}[c]{@{}l@{}}mixing in\\ different ways\end{tabular}} \\ \cline{2-3}
		&
		\begin{tabular}[c]{@{}l@{}} tell me a kitchen activity related verb which is done while stirring a \\ big bowl containing brownie mixture which is not transitive can be used \\ in the form something, shows achievement, requires medium motion, \\ requires time in order of seconds, is performed in solitary, which \\ requires arms to be used for the action but not head, legs, torso or any\\ other body parts, changes the external world and which moves the object\end{tabular} &
		blend &
		\\ \cline{2-3}
		&
		\begin{tabular}[c]{@{}l@{}} tell me a kitchen activity related verb which is done while stirring a \\ big bowl containing brownie mixture which is not transitive can be used \\ in the form something, shows achievement, requires low motion, requires \\ time in order of seconds, is performed in solitary, which requires arms \\ to be used for the action but not head, legs, torso or any other body \\ parts and changes the state of the object \end{tabular} &
		fold &
		\\ \cline{2-3}
		&
		\begin{tabular}[c]{@{}l@{}} tell me a kitchen activity related verb which is done while stirring a \\ big bowl containing brownie mixture which is not transitive can be used \\ in the form something, shows achievement, requires medium motion, \\ requires time in order of seconds, is performed in solitary, which \\ requires arms to be used for the action but not head, legs, torso or any\\ other body parts, which moves the object and changes the state of the \\ object \end{tabular} &
		incorporate &
		\\ \hline
  \multirow{2}{*}{\begin{tabular}[c]{@{}l@{}}pour oil into\\ small measuring cup\end{tabular}} &
		\begin{tabular}[c]{@{}l@{}} tell me a kitchen activity related verb which is done while pouring oil \\ into a measuring cup which can be used in the form something, shows an \\ achievement, requires medium motion, requires time in order of seconds, \\ is performed in solitary, which requires arms to be used for the action \\ but not head, legs, torso or any other body parts, moves the object \\ somewhere, and changes the state of the object \end{tabular} &
		dispense &
		\begin{tabular}[c]{@{}l@{}}read measuring\\ cup labels\end{tabular} \\ \cline{2-4} 
		&
		\begin{tabular}[c]{@{}l@{}} tell me a kitchen activity related verb which is done while pouring oil \\ into a measuring cup which can be used in the form something, shows an \\ achievement, requires medium motion, requires time in order of seconds, \\ is performed in solitary, which requires arms to be used for the action \\ but not head, legs, torso or any other body parts, moves the object \\ somewhere, and changes the state of the object \end{tabular} &
		transfer &
		pour in \\ \hline
	\end{tabular}
\end{table}
\section{Discussion and Future Directions}
\label{discussion}
Undoubtedly, the development of \method{} opens up a greater avenue for further investigation and analysis. We envision that in the near future, the following key ideas may galvanize the research in this direction to develop more explainable and sophisticated modeling approaches surrounding HAR. The details follow.
\subsection{Making \method{} Multimodal}
The current version of \method{} uses an unimodal sensing approach, only extracting information from the wearable locomotive sensors. Interestingly, smart environments usually capture data from many different modalities. Together, these modalities can enhance the performance and applicability of \method. For example, in the HSD dataset (Section~\ref{datasets}), the smart home was also equipped with sensors attached to furniture. Indeed, the output of these sensors can be used to first impute the coarsely labeled dataset with actions recognized from some atomic activity like \say{opening cupboard}. Subsequently, this imputed dataset can be used directly with \method{} to identify the remaining micro-activities further. Similarly, approaches like~\cite{acconotate,chatterjee2020laso} can also be used to add additional labels to the data, which can then be used with \method{} to provide a more explainable decomposition of the activities occurring in the smart environment.
\subsection{Bringing LLMs for Virtual Supervision}
Obtaining information from the web for sensor annotation has been an idea that has been well explored before as well. For example,~\cite{alirezaie2013automatic} used \textit{abductive reasoning} to annotate medical grade sensors. Recently, with the development and availability of LLMs, there has been a significant rise in works like~\cite{liu2023large,silva2023gpt4,10.1145/3581791.3597366}. A potential approach in this direction can also be made by utilizing general-purpose latent representations like verb attribute embeddings to convert predictions into queries. This can then be used to validate the micro-activities identified by \method{} using LLMs as a source for added supervision.
\section{Conclusion}
\label{conclu}
Complex human activities are often sequences of different short-duration micro-activities. Naturally, a system for recognizing and modeling these sequences will allow the development of more sophisticated and explainable HAR models that could be used for different intelligent services. Following this motivation, we, in this paper, developed the framework \method{}, which can identify the hidden micro-activities present within a broad macro-activity without requiring external supervision or access to granular annotations. Notably, this top-down approach allows \method{} to be more realistic and cost-effective than the conventional bottom-up approaches, which heavily depend on fine-grained annotated data to identify the macro-activity. To identify the hidden micro-activities in a bottom-up manner, \method{} first detects the micro-activity boundaries using unsupervised change-point detection and identifies the micro-activities from the detected patterns using a generalized zero-shot approach. A carefully designed amalgamation of these two approaches allows \method{} to be robust, unsupervised, label-efficient, and explainable.

\bibliographystyle{ACM-Reference-Format}
\bibliography{ref/ref}

\end{document}